\documentclass[fleqn,usenatbib]{mnras}

\usepackage{newtxtext,newtxmath}
\usepackage[T1]{fontenc}
\usepackage{ae,aecompl}

\usepackage{graphicx}	% Including figure files
\usepackage{amsmath}	% Advanced maths commands
\usepackage{amssymb}	% Extra maths symbols
\usepackage{bm}
\usepackage{tikz}
\usepackage{subcaption}
\usetikzlibrary{arrows,shapes,positioning}

\newcommand{\ttt}[1]{\ensuremath{^{(#1)} }}
\newcommand{\nnn}[1]{\ensuremath{{[#1]} }}
\title[CMU DeepLens]{CMU DeepLens: Deep Learning For Automatic Image-based Galaxy-Galaxy Strong Lens Finding}

\author[Lanusse et al.]{Fran\c{c}ois Lanusse,$^{1}$\thanks{E-mail: flanusse@andrew.cmu.edu}
Quanbin Ma,$^{2}$
Nan Li,$^{3,4}$
Thomas E. Collett,$^5$
Chun-Liang Li,$^2$
\newauthor
Siamak Ravanbakhsh,$^{2}$
Rachel Mandelbaum$^{1}$
and Barnab\'as P\'oczos$^{2}$
\\
$^{1}$McWilliams Center for Cosmology, Department of Physics, 
Carnegie Mellon University, Pittsburgh, PA 15213, USA\\
$^{2}$School of Computer Science, Carnegie Mellon University, Pittsburgh, PA 15213, USA\\
$^{3}$High Energy Physics Division, Argonne National Laboratory, Lemont, IL 60439, USA\\
$^{4}$Department of Astronomy \& Astrophysics, The University
of Chicago, 5640 South Ellis Avenue, Chicago, IL 60637, USA\\
$^{5}$Institute of Cosmology and Gravitation, University of Portsmouth, Burnaby Rd, Portsmouth, PO1 3FX, UK
}

\date{Accepted XXX. Received YYY; in original form ZZZ}

\pubyear{2017}

\begin{document}
\label{firstpage}
\pagerange{\pageref{firstpage}--\pageref{lastpage}}
\maketitle

\begin{abstract}
Galaxy-scale strong gravitational lensing is not only a valuable probe of the dark matter distribution of massive galaxies, but can also provide valuable cosmological constraints, either by studying the population of strong lenses or by measuring time delays in lensed quasars. Due to the rarity of galaxy-scale strongly lensed systems, fast and reliable automated lens finding methods will be essential in the era of large surveys such as LSST, Euclid, and WFIRST. To tackle this challenge, we introduce \texttt{CMU DeepLens}, a new fully automated galaxy-galaxy lens finding method based on Deep Learning. This supervised machine learning approach does not require any tuning after the training step which only requires realistic image simulations of strongly lensed systems. We train and validate our model on a set of  20,000 LSST-like mock observations including a range of lensed systems of various sizes and signal-to-noise ratios (S/N). We find on our simulated data set that for a rejection rate of non-lenses of 99\%, a completeness of 90\% can be achieved for lenses with Einstein radii larger than 1.4\arcsec\ and S/N larger than 20 on individual $g$-band LSST exposures. Finally, we emphasize the importance of realistically complex simulations for training such machine learning methods by demonstrating that the  performance of models of significantly different complexities cannot be distinguished on simpler simulations.
We make our code publicly available at \url{https://github.com/McWilliamsCenter/CMUDeepLens}.
\end{abstract}


%%%%%%%%%%%%%%%%%%%%%%%%%%%%%%%%%%%%%%%%%%%%%%%%%%

%%%%%%%%%%%%%%%%% BODY OF PAPER %%%%%%%%%%%%%%%%%%

\section{Introduction}

Strong gravitational lensing, the serendipitous appearance of multiple images or extended arcs due to distant quasars or galaxies almost directly behind a massive object along the same line-of-sight, finds many important applications both  astrophysical and cosmological. It is in particular a well-established probe of overall gravitational potential of massive galaxies. For example, moderate-sized samples of galaxy-scale strong lenses from the Sloan Lens ACS (SLACS) survey \citep{Bolton2006} have been used to derive ensemble constraints on the total matter profile in massive elliptical galaxies \citep[e.g.][]{Koopmans2006, Auger2010a, Barnabe2011}.  As a cosmological probe, time delays in multiply imaged strongly lensed quasars can be used to derive independent constraints on the Hubble constant $H_0$ \citep[e.g.][]{Refsdal1964, Suyu2010, Bonvin2017}, but even without time delays strongly lensed systems can constrain cosmological parameters and in particular the dark energy equation of state \citep[e.g.][]{Collett2014, Cao2015}. See, e.g., \citet{Treu2010} for a review of results using strong lensing by galaxies.

Upcoming large sky surveys such as the Large Synoptic Survey Telescope (LSST\footnote{\url{https://www.lsst.org/lsst/}}; \citealt{LSST2009}), Euclid\footnote{\url{http://sci.esa.int/euclid/}, \url{http://www.euclid-ec.org}} 
\citep{Laureijs2011}, and the Wide-Field Infrared Survey Telescope (WFIRST\footnote{\url{https://wfirst.gsfc.nasa.gov/}}; \citealt{Spergel2015}) will produce datasets that include unprecedented numbers of galaxy-scale strong lenses.  For example, \citet{Collett2015} predicts $>10^5$ galaxy-scale strong lens systems in LSST and Euclid.  This lens population is expected to be dominated by intermediate-redshift ($z\sim 0.5$--1) elliptical galaxies, with blue source galaxies. Clearly, the challenge will be to detect those galaxy-scale strong lenses efficiently and in a way that results in an easily-quantifiable selection function for further scientific investigation.  LSST and Euclid present different challenges for automated galaxy-scale strong lens detection. The LSST imaging will have 6 photometric bands, enabling us to distinguish between objects with different colors, but at low resolution compared to space-based imaging.  In contrast, the Euclid imaging will be far higher resolution but with only a single optical passband. Thus, neither survey achieves the optimal setup for galaxy-scale strong lens detection, i.e., high-resolution multi-color imaging.  The space-based, multi-band near-infrared data from WFIRST in principle provides a dataset that is closer to optimal, though the galaxy populations probed may differ due to the fact that the imaging is at near-infrared wavelengths.

Most automated image-based lens finding methods proposed so far have been based on the detection of arc and ring structures in the images. The  \texttt{ARCFINDER} method \citep{Alard2006, More2012} relies for instance on a local elongation estimator, at the pixel level.  Other approaches to arc finding purely based on morphology include those described by \cite{Seidel2007, Kubo2008, Bom2016}.  
When considering specifically galaxy-scale strong lenses, arcs can often be obscured by the light of the lens galaxy. Several methods have  been proposed to perform a first subtraction step of the lens galaxy to facilitate the detection of faint arcs. The \texttt{RingFinder} algorithm proposed in \cite{Gavazzi2014} uses a multi-band differencing scheme to reveal faint blue arcs surrounding early type galaxies. \cite{Joseph2014} proposed an alternative subtraction scheme, which does not require multi-band images, based on a PCA decomposition of the lens light profile.
A different approach based on physical lens modeling was initially proposed in \cite{Marshall2009} for high resolution space-based images and revisited in \cite{Brault2015} for ground-based images. In this class of methods, a simple model including a background source and a foreground deflector is fitted on each lens candidate, and a classification as a possible lens is made based on the predicted model parameters. 

As an example of automated lens searches, some of these methods were applied to the Strong Lensing Legacy Survey \citep[SL2S;][]{Cabanac2007}, a survey from the Canada-France-Hawaii Telescope Legacy Survey (CFHTLS) covering an approximate area of 150 deg$^2$. The \texttt{ARCFINDER} was  applied to the survey in \cite{More2012}, producing a final sample of 127 potential lenses, mostly on group and cluster mass scales. The authors report that the  algorithm produced an average of 1,000 candidates per deg$^2$ requiring a significant amount of further visual inspection. The \texttt{RingFinder} algorithm was also applied to the CFHTLS data \citep{Gavazzi2014} but concentrated on galaxy-scale lenses by pre-selecting early-type galaxies. This algorithm produced on average a more manageable 18 candidates per deg$^2$, and a final sample of 330 high probability lenses after visual inspection. In this particular case, the authors reported that visual inspection required an estimated 30 person-minutes per deg$^2$. At this rate, next generation surveys such as Euclid ($\sim$15000 deg$^2$) or LSST ($\sim$20000 deg$^2$) would still require a very significant investment of human time. 

One approach to scale the visual inspection effort to the size of  these surveys is to use crowdsourcing. This is the idea behind the \texttt{Space Warps} project \citep{Marshall2015, More2015},  which crowdsourced the visual inspection of a sample of 430,000 images from the  CHFTLS to a crowd of 37,000 citizen scientists, yielding a new sample of gravitational lens candidates. 59 of these candidates were previously missed by robotic searches, while  20 known lenses were missed by the volunteers. These results show a complementarity between the two approaches, and crowdsourced visual inspection could prove very valuable in screening lens candidates found by automated searches. The authors further estimate that a similar crowdsourcing effort can be scaled up to LSST sizes, where a considerable crowd of $10^6$ volunteers could visually inspect $10^6$ LSST targets in a matter of weeks. 

The image classification problem involved in strong lens finding is a notoriously challenging task that has received considerable attention in the broader field of computer vision and machine learning. Very recently, a new class of models based on the \emph{Deep Learning} framework \citep{LeCun2015} has been able to surpass human accuracy in similar image classification tasks \citep{He2015}. Such models are therefore extremely promising for gravitational lens searches as they could prove more reliable than non-expert human inspection and therefore dramatically reduce the amount of human time investment for future surveys. Some Deep Learning models have already been proposed in an astrophysical context, most notably for automatic identification of galaxy morphologies \citep{Dieleman2015}, performing star-galaxy classification \citep{Kim2017}, estimating photometric redshifts \citep{Hoyle2016}, or for generative models of galaxy images \citep{Ravanbakhsh2016}.

The idea of using Deep Learning for detecting strongly lensed systems has recently been explored by several groups. For instance, \cite{Petrillo2017} recently published an application on the  Kilo Degree Survey \citep[KiDS;][]{DeJong2015}. This method, as well as all other Deep Learning methods currently under investigation for this task are based on deep \emph{Convolutional Neural Networks} \citep[CNNs;][]{Lecun1998, Krizhevsky2012}, a powerful architecture for image detection and classification tasks.

In this paper, we present a new approach to strong lens finding, based on deep \emph{Residual Networks} \citep{He2015}, an advanced variation of CNNs that constitutes the current state of the art in image classification. As a result, our proposed architecture, named \texttt{CMU DeepLens}, or \texttt{DeepLens} in short, has recently been found to outperform most CNN-based lens finders in a recent blind challenge organised by the Euclid strong lensing working group\footnote{\url{http://metcalf1.bo.astro.it/blf-portal/gg_challenge.html}} (Metcalf et al. 2017, in prep.). 

Contrary to most previous methods, such a supervised machine learning approach does not make any prior assumptions on specific features or physical models and instead lets the machine learn from the provided training data which features are the most relevant to the detection of strong lenses. In addition to characterising the performance of our baseline architecture using simulations of lens systems in LSST, we also explore the impact of model and simulation complexity on machine learning-based lens finders.

This paper is structured as follows. In \autoref{sec:deep_learning}, we provide some background on the Deep Learning framework and introduce the building blocks used in our classifier. In \autoref{sec:deeplens}, we detail the \texttt{DeepLens} architecture itself as well as the training procedure. We then describe in \autoref{sec:sims} the strong lensing simulation pipeline that was used to generate a set of training and validation images. These simulations are used in \autoref{sec:results} to quantify the performance of our model. Finally,  \autoref{sec:discussion} discusses current limitations as well as several avenues for further improvement of our results, with conclusions provided in \autoref{sec:conclusion}. 

\section{Deep Learning Background}
\label{sec:deep_learning}

In this first section, we provide a brief overview of the Deep Learning framework and introduce the specific components used in our model.

\subsection{The Deep Learning Revolution}
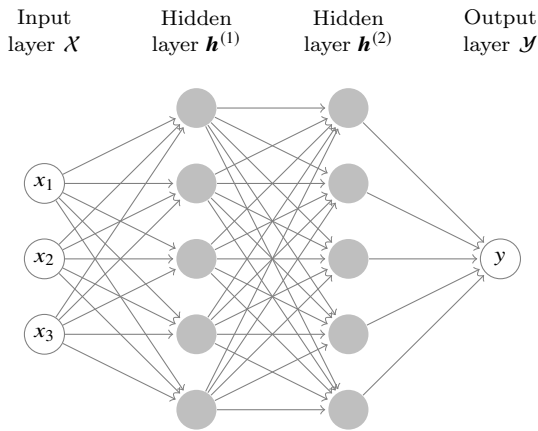
\begin{figure}
	\centering
	\def\layersep{2.0cm}
\begin{tikzpicture}[shorten >=1pt,->,draw=black!50, node distance=\layersep]
    \tikzstyle{every pin edge}=[<-,shorten <=1pt]
    \tikzstyle{neuron}=[circle,fill=black!25,minimum size=15pt,inner sep=0pt]
    \tikzstyle{input neuron}=[neuron, fill=white,draw];
    \tikzstyle{output neuron}=[neuron, fill=white, draw];
    \tikzstyle{hidden neuron}=[neuron, fill=gray!50];
    \tikzstyle{annot} = [text width=4em, text centered]

    % Draw the input layer nodes
    \foreach \name / \y in {1,...,3}
    % This is the same as writing \foreach \name / \y in {1/1,2/2,3/3,4/4}
        \node[input neuron] (I-\name) at (0,-\y-0.8) {$x_{\y}$};

    % Draw the hidden layer nodes
    \foreach \name / \y in {1,...,5}
        \path[yshift=0.2cm]
            node[hidden neuron] (H-\name) at (\layersep,-\y cm) {};
            
    % Draw the hidden layer nodes
    \foreach \name / \y in {1,...,5}
        \path[yshift=0.2cm]
            node[hidden neuron] (H2-\name) at (2*\layersep,-\y cm) {};

    % Draw the output layer node
    \node[output neuron, right of=H2-3] (O) {$y$};

    % Connect every node in the input layer with every node in the
    % hidden layer.
    \foreach \source in {1,...,3}
        \foreach \dest in {1,...,5}
            \path (I-\source) edge (H-\dest);
            
    % Connect every node in the input layer with every node in the
    % hidden layer.
    \foreach \source in {1,...,5}
        \foreach \dest in {1,...,5}
            \path (H-\source) edge (H2-\dest);

    % Connect every node in the hidden layer with the output layer
    \foreach \source in {1,...,5}
        \path (H2-\source) edge (O);

    % Annotate the layers
    \node[annot,above of=H-1, node distance=1cm] (hl) {Hidden layer $\bm{h}^{(1)}$};
    \node[annot,above of=H2-1, node distance=1cm] (hl2) {Hidden layer $\bm{h}^{(2)}$};
    \node[annot,left of=hl] {Input layer $\mathcal{X}$};
    \node[annot,right of=hl2] {Output layer $\mathcal{Y}$};
\end{tikzpicture}
	\caption{Conventional multilayer perceptron, a feedforward neural network with fully connected hidden layers. Each arrow represents a directed weighted connection.}
	\label{fig:nn}
\end{figure}

Artificial Neural Networks (ANNs) have been an established tool for classification and regression tasks for several decades. In fully connected feedforward models such as the Multilayer Perceptron (MLP), each layer is composed of elementary units (the neurons) performing a simple weighted linear combination over the outputs of all units in the previous layer, followed by the application of a non-linear transform, also known as the \emph{activation function}. This architecture is illustrated in \autoref{fig:nn}. Let $1 \leq \ell \leq L$ denote the neural layer in a deep architecture, and let $N_\ell$ denote the number of outputs of layer $\bm{h}^{(\ell)}$.
Using tensor notations, the output of a given layer $\bm{h}\ttt{\ell} \in \Re^{N_\ell}$ of the network can be expressed in terms of the output of the previous layer $\bm{h}\ttt{\ell-1}$ according to:
\begin{equation}
	\bm{h}\ttt{\ell} = f( \mathbf{W}\ttt{\ell} \cdot \bm{h}\ttt{\ell-1} + \bm{b}\ttt{\ell})
\end{equation}
where $f: \Re^{N_{\ell}} \to \Re^{N_{\ell}}$ is the element-wise activation function, such as the sigmoid-shaped logistic function $f(\bm{h}) =  1/(1 + \exp(-\bm{h}))$, $\mathbf{W}\ttt{\ell} \in \Re^{N_{\ell} \times N_{\ell-1}}$ is a dense weight matrix, and $\bm{b}\ttt{\ell} \in \Re^{N_{\ell}}$ is a vector of additive biases applied before the activation function.
Here the input $\bm{x} = \bm{h}\ttt{0}$ is identified by $\ell = 0$ and output $\bm{y} = \bm{h}\ttt{L}$ is the final layer.

Such a model is trained to perform a specific task by optimising its parameters $\{\mathbf{W},\bm{b}\}$ as to minimise a given \textit{loss function}. This optimisation is  performed by a Stochastic Gradient Descent (SGD) algorithm which iteratively updates the model weights by taking small gradient steps computed over a randomly selected sub-sample of the training set. The computation of these gradients are made tractable in practice using the \emph{backpropagation} algorithm \citep{Rumelhart1986}, which simply applies the chain rule to efficiently obtain the derivative of the the loss function with respect to the model parameters. 
This, relies on the idea that the gradients at layer $\ell$ can efficiently be computed by using the gradients at layer ${\ell + 1}$. 
A crucial point of this training procedure is that the gradients of the model are computed starting from the last hidden layer $\ell = L$ and then propagated through the network back to the first layer $\ell = 1$. 

If trained well, the performance of such a feedforward neural network is expected to increase with the depth of the model, as additional layers allow for a more complex mapping from the input to the output. Even using one hidden layer, a feedforward network --with sufficient number of hidden units-- is known to be a universal approximator~\citep{Hornik1989} -- that is it can approximate any function to arbitrary precision.

However, until very recently, deep networks with many hidden layers had remained completely unpractical as they were notoriously difficult to train. The main reason is the so-called \textit{vanishing gradient} effect where the gradient of the cost function decreases exponentially due to the repeated chain rules and eventually vanishes as it is backpropagated through the layers during training. As a result, parameters in lower-level layers could not be tuned well in practice, thus greatly limiting the depth of typical models to a few hidden layers.

While this limitation had remained unsolved for several decades, the Deep Learning revolution was brought about by a conjunction of factors: the emergence of effective procedures to train deep architectures, the explosion of the volume of available training data, and the increase in computing power through Graphical Processing Units (GPU). Amongst the innovations that allowed for deeper models, the ReLU (for rectified linear unit) activation function was introduced in \cite{Nair2010}. This simple function, defined as $f(x) = \max(x, 0)$, does not saturate contrary to conventional sigmoid functions, leading to much stronger gradients which can be propagated deeper in the models during training. The availability of much larger training sets as well as new regularisation techniques (e.g. dropout regularisation; \citealt{Hinton2012, Srivastava2014}) have further made it possible to train large networks without over-fitting. Finally, efficient implementations of neural network architectures on GPUs has accelerated the training process by several orders of magnitudes, compared to CPU implementations. Combining these factors and innovations with the pre-existing Convolutional Neural Network architecture \citep[][see the next section for details]{Lecun1998} deep neural networks have suddenly been able to reach significant depth and achieve state-of-the-art results in image classification problems \citep{Krizhevsky2012}.

Deep architectures now achieve state-of-the-art (and some times superhuman) performance in a wide range of applications in computer vision, natural language processing, bioinformatics, data-mining and computer games. We refer interested readers to \cite{Goodfellow2016} for a general introduction.

\subsection{Convolutional Neural Networks}

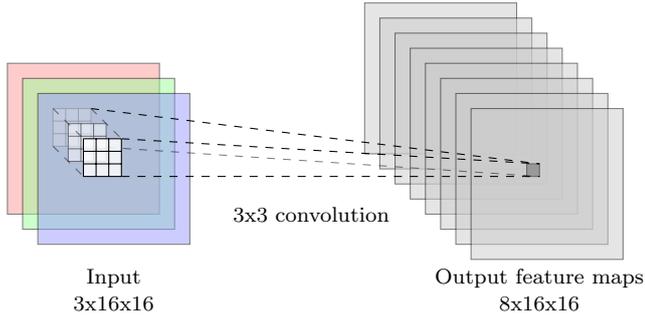
\begin{figure}
	\centering
	
	\begin{tikzpicture}
	\def\layersep{0.2cm}
	
	\newcommand{\mygrid}{\tikz{\draw[step=1.6666mm,opacity=0.5] (0,0)  grid (0.5,0.5);}}
	
	% Draw the frames  
	\filldraw[fill=red!40!white, draw=black, opacity=0.5] (0.,2.) rectangle (2.,0);
	\filldraw[fill=green!40!white, draw=black, opacity=0.5] (0.2,1.8) rectangle (2.2,-0.2);
	\filldraw[fill=blue!40!white, draw=black, opacity=0.5] (0.4,1.6) rectangle (2.4,-0.4);
	
	% Draw the next frames
    \foreach \y in {1,...,8}
		\filldraw[fill=black!20!white,opacity=0.5]  (4.5+\y/5,3 - \y/5 ) rectangle (6.5+\y/5,1 - \y/5);

	% Draw the convolution
	\draw[dashed] (1.5,1) -- (7.,0.6666);
	\draw[dashed] (1.1,1.4) -- (7,0.6666);
	
	\draw[dashed] (1.5,0.5) -- (7.,0.5);
	\draw[dashed, opacity=0.5] (1.1,0.9) -- (7.,0.5);
	
	\filldraw[fill=black!60!white, draw=black, opacity=0.5] (6.83334, 0.5) rectangle (7.,0.6666);

	\draw (0.85,1.15) node[rectangle, inner sep=0, opacity=0.2, fill=white] {\tikz{\draw[step=1.6666mm,opacity=0.2] (0,0)  grid (0.5,0.5);}};
	
	\draw (1.05,0.95) node[rectangle, inner sep=0, opacity=0.5, fill=white] {\tikz{\draw[step=1.6666mm,opacity=0.5] (0,0)  grid (0.5,0.5);}};
	\draw (1.25,0.75) node[rectangle, inner sep=0, opacity=0.8, fill=white] {\tikz{\draw[step=1.6666mm,opacity=1] (0,0)  grid (0.5,0.5);} };
	
	\draw[dashed, opacity=0.5] (1.1,1.4) -- (1.5,1);
	\draw[dashed, opacity=0.5] (0.6,1.4) -- (1.,1);
	\draw[dashed, opacity=0.5] (0.6,0.9) -- (1.,0.5);
	
	%Add text
	\draw[] (1.4, -1) node[align=center]{Input\\3x16x16};
	\draw[] (7, -1) node[align=center]{Output feature maps\\8x16x16};
	\draw[] (4, -0.) node[align=center]{3x3 convolution};

	\end{tikzpicture}
	\caption{Illustration of one convolutional layer, applying $3\times3$ convolution kernels on an RGB image to produce a set of eight feature maps. Only the convolution by the first kernel is illustrated for clarity; each feature map is obtained by convolving the input image with a different convolution kernel.}
	\label{fig:convnet}
\end{figure}

Taking into account the specific topological structure and properties of natural images, \cite{Lecun1998} introduced the Convolutional Neural Network (CNN) as an efficient alternative to fully connected architectures for image processing. The building block of CNNs is the \textit{convolutional layer} which outputs a set of \textit{feature maps} by convolving an input image with learned local filters. This process is illustrated in \autoref{fig:convnet} where an input RGB image is convolved with a set of eight 3x3 filters to yield eight output feature maps.

More formally, let us consider $\bm{h}^{(\ell -1)}$ the input of convolution layer $\ell$. Considering square images for simplicity, we will denote the height and width in pixels of  $\bm{h}^{(\ell -1)}$ by $N_{\ell -1}$, while its \emph{depth} (i.e. the number of bands or feature maps) will be denoted by $K_{\ell - 1}$, so that  $\bm{h}^{(\ell -1)} \in \Re^{K_{\ell-1} \times N_{\ell-1} \times N_{\ell-1}}$. In a similar fashion, we will note $\bm{h}^{(\ell)} \in \Re^{K_{\ell} \times N_{\ell} \times N_{\ell}}$ the output of convolution layer $\ell$. In the case of the illustration of \autoref{fig:convnet}, $N_{\ell -1} = N_{\ell} = 16$ while $K_{\ell -1}= 3$ for the input RGB image and $K_{\ell}=8$ for the output feature-maps. With these notations, a single output feature-map of $\bm{h}^{(\ell)}$ for $1 \leq k \leq K_\ell$ is expressed as:
\begin{equation}
	\bm{h}_{k}\ttt{\ell} = f \left( \sum_{k'} \mathbf{W}_{k',k}\ttt{\ell} \ast \bm{h}_{k'}\ttt{\ell-1} + b\ttt{\ell}_{k} \right) \;, \label{eq:convnet}
\end{equation}
where $\mathbf{W}^{(\ell)} \in \Re^{K_{\ell-1} \times K_{\ell} \times I_{\ell} \times I_{\ell}}$ is referred to as the \emph{convolution kernel}, and it contains a different filter of size $I_{\ell} \times I_{\ell}$ for each combination of input and output channels. As in the fully-connected architecture, $b\ttt{\ell}_{k} \in \Re$ is a bias parameter (i.e., one scalar parameter per output channel) and $f$ is a nonlinearity. 

The size of the convolution kernel is generally limited to a few pixels (e.g., $I_{\ell} = 3$). To capture larger scale information, CNNs follow a hierarchical multi-scale approach by interleaving convolution layers with so-called \emph{pooling layers}, which apply a downsampling operation to the  feature maps. A CNN architecture is a stack of convolution layers and pooling layers, converting the input image into an increasing number of feature maps of progressively coarser resolution. The final feature maps $\bm{h}\ttt{L}$ can capture information on the scale of the input image and can reach a high-level of abstraction.
Combined with efficient GPU implementations, CNNs have become a standard model for image detection and classification tasks.

\subsection{Deep Residual Networks}

As mentioned in the previous section, the performance of a deep network generally increases with the number of layers, up to the point at which the model becomes too difficult to train and performances start to degrade. This general rule still applies to CNN architectures, which despite reaching greater depths than previous models have been limited to around a dozen layers. To overcome this well-known difficulty, machine learning research has recently been focused on developing alternative architectures beyond simple CNNs, in an attempt to build deeper models that can still be efficiently trained.

Several very recent developments have lead to significant improvements in model accuracy for classification and detection tasks. For instance, the \textit{Inception} architecture \citep{Szegedy2015} allowed the GoogLeNet model to reach a depth of 22 layers, and as a result to win the ImageNet Large-Scale Visual Recognition Challenge 2014 (ILSVRC2014), one of the benchmarks in image classification and object detection. Even more recently, \citet{He2015} introduced the \textit{deep residual learning} framework, which allowed the authors to increase the depth of their convolutional model by one order of magnitude to 152 layers while still improving detection and classification accuracy and ultimately winning the ILSVRC2015. These  results were  even further improved in a follow-up work \citep{He2016} where the authors successfully trained models as deep as 1000 layers while still improving classification accuracy.  The  model proposed in our work is directly based on this state-of-the-art deep residual network, or \emph{resnet}, architecture.

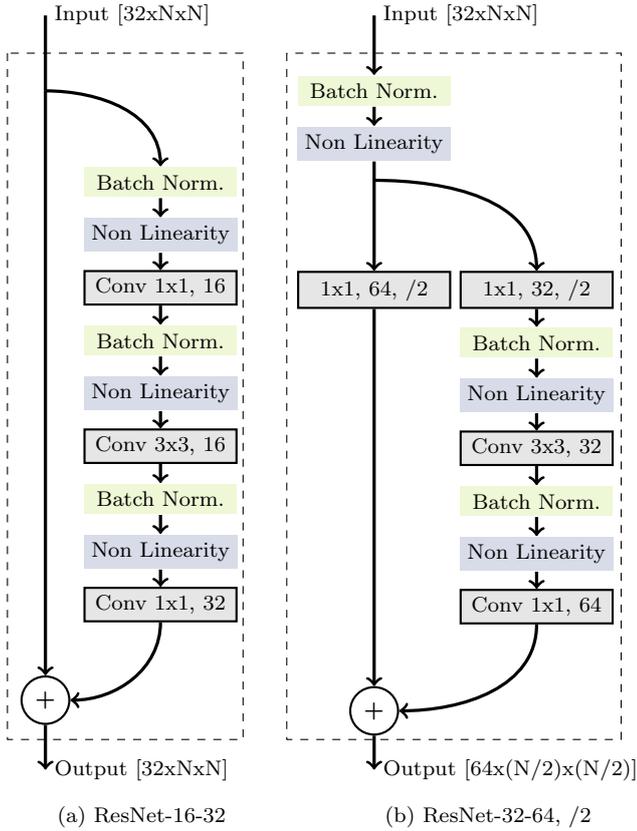
\begin{figure}
\centering
\definecolor{mycolor1}{HTML}{aadb32}
\definecolor{mycolor2}{HTML}{5bc862}
\definecolor{mycolor3}{HTML}{27ad80}
\definecolor{mycolor4}{HTML}{208f8c}
\definecolor{mycolor5}{HTML}{2c718e}
\definecolor{mycolor6}{HTML}{3b518a}
\definecolor{mycolor7}{HTML}{472b7a}
\begin{subfigure}[t]{0.2\textwidth}
	\begin{tikzpicture}[  
	block/.style    = {draw, thick, rectangle, minimum height = 1.25em, minimum width = 2cm, node distance = 0.7cm, fill=black!10!white},
	relu/.style    = { thick, rectangle, minimum height = 1.25em, minimum width = 2cm, node distance = 0.7cm, fill=mycolor6!20!white},
	batch/.style    = { thick, rectangle, minimum height = 1.25em, minimum width = 2cm, node distance = 0.7cm, fill=mycolor1!20!white},
    sum/.style      = {draw, circle, node distance = 1.cm}, % Adder
	input/.style    = {coordinate, node distance = 1.cm}, % Input
  	output/.style   = {coordinate, node distance = 1.cm} % Output
]

%Nodes
\node[input] (input1)  {};
\node[batch,below right=1.cm and 0.5cm of input1] (batch_norm1) {Batch Norm.};
\node[relu, below of=batch_norm1] (relu1) {Non Linearity};
\node[block, below of=relu1]  (layer1)   {Conv 1x1, 16};
\node[batch, below of=layer1 ] (batch_norm2) {Batch Norm.};
\node[relu, below of=batch_norm2] (relu) {Non Linearity};
\node[block, below of=relu]  (layer2)   {Conv 3x3, 16};
\node[batch, below of=layer2 ] (batch_norm3) {Batch Norm.};
\node[relu, below of=batch_norm3] (relu3) {Non Linearity};
\node[block, below of=relu3]  (layer3)   {Conv 1x1, 32};

\node[sum, below = 7.75cm of input1, thick] (suma) {\textbf{+}};
\node[output, below of=suma] (output1)  {};
 
%Lines
\draw[->, very thick] (input1) --  (suma);
\draw[->, very thick] (input1) to [out=0, in=90]  (batch_norm1.north);
\draw[->, very thick] (batch_norm1) --  (relu1);
\draw[->, very thick] (relu1) --  (layer1);
\draw[->, very thick] (layer1) --  (batch_norm2);
\draw[->, very thick] (batch_norm2) --  (relu);
\draw[->, very thick] (relu) --  (layer2);
\draw[->, very thick]  (layer2) -- (batch_norm3);
\draw[->, very thick] (batch_norm3) --  (relu3);
\draw[->, very thick] (relu3) --  (layer3);
\draw[->, very thick] (layer3.south) to [out=-90, in=0 ]  (suma.east);

% Draw bounding box of the resnet unit
\draw[dashed] ( -0.5,0.5) rectangle (2.6,-8.6);
\draw[-, very thick] (0,1) node[right]{Input [32xNxN]} -- (input1);
\draw[-> , very thick] (suma) -- (0, -9) node[right]{Output [32xNxN]};

\end{tikzpicture}
\caption{ResNet-16-32}
\label{fig:resnet_unit}
\end{subfigure}~
\begin{subfigure}[t]{0.3\textwidth}
	\begin{tikzpicture}[  
	block/.style    = {draw, thick, rectangle, minimum height = 1.25em, minimum width = 2cm, node distance = 0.7cm, fill=black!10!white},
	relu/.style    = { thick, rectangle, minimum height = 1.25em, minimum width = 2cm, node distance = 0.7cm, fill=mycolor6!20!white},
	batch/.style    = { thick, rectangle, minimum height = 1.25em, minimum width = 2cm, node distance = 0.7cm, fill=mycolor1!20!white},
    sum/.style      = {draw, circle, node distance = 1.cm}, % Adder
	input/.style    = {coordinate, node distance = 1.cm}, % Input
  	output/.style   = {coordinate, node distance = 1.cm} % Output
]
%Nodes

\node[batch] (batch_norm1) {Batch Norm.};
\node[relu, below of=batch_norm1] (relu1) {Non Linearity};

\node[block, below right=1.45cm and 0.1cm of relu1]  (layer1)   {1x1, 32, /2};
\node[block, below = 1.45cm of relu1]  (proj1)   {1x1, 64, /2};
\node[batch, below of=layer1 ] (batch_norm2) {Batch Norm.};
\node[relu, below of=batch_norm2] (relu) {Non Linearity};
\node[block, below of=relu]  (layer2)   {Conv 3x3, 32};
\node[batch, below of=layer2 ] (batch_norm3) {Batch Norm.};
\node[relu, below of=batch_norm3] (relu3) {Non Linearity};
\node[block, below of=relu3]  (layer3)   {Conv 1x1, 64};
\node[sum, below = 5.cm of proj1, thick] (suma) {\textbf{+}};
\node[output, below of=suma] (output1)  {};
 
%Lines
\draw[->, very thick] (proj1) --  (suma);
\draw[->, very thick] (relu1.south) ++ (0,-0.25) to [out=0, in=90]  (layer1.north);
\draw[->, very thick] (batch_norm1) --  (relu1);
\draw[->, very thick] (relu1) --  (proj1);
\draw[->, very thick] (layer1) --  (batch_norm2);
\draw[->, very thick] (batch_norm2) --  (relu);
\draw[->, very thick] (relu) --  (layer2);
\draw[->, very thick]  (layer2) -- (batch_norm3);
\draw[->, very thick] (batch_norm3) --  (relu3);
\draw[->, very thick] (relu3) --  (layer3);
\draw[->, very thick] (layer3.south) to [out=-90, in=0 ]  (suma.east);

% Draw bounding box of the resnet unit
\draw[dashed] ( -1.15,0.5) rectangle (3.25,-8.6);
\draw[->, very thick] (0,1) node[right]{Input [32xNxN]} -- (batch_norm1);
\draw[->, very thick] (suma) -- (0, -9) node[right]{Output [64x(N/2)x(N/2)]};
\end{tikzpicture}
\caption{ResNet-32-64, /2}
\label{fig:resnet_unit_dec}
\end{subfigure}
	\caption{Pre-activated bottleneck residual units, the building blocks of a residual network architecture. Left (a): Undecimated ResNet-16-32 unit, preserving the size and depth of the input. Right (b):  ResNet-32-64,/2 unit simultaneously increasing the depth of the output (from 32 channels to 64) and downsampling by a factor 2 its resolution.}
	\label{fig:residual_block}
\end{figure}

\emph{Residual learning} aims to tackle the problem of vanishing gradients and difficult optimisation of deep networks by introducing so-called \emph{shortcut} connections between the input and output of a stack of a few convolution layers, so that instead of learning how to map the input to the output, the convolution layers are only learning their difference, hence the term residual learning. In other words, instead of learning directly a given non linear mapping $\mathcal{H}(\bm{x})$ between input and outputs, residual networks are trained to learn the residual mapping with respect to the identity: $\mathcal{F}(\bm{x}) = \mathcal{H}(\bm{x})  - \bm{x}$. An illustration of a simple residual unit proposed in \cite{He2016} is provided on \autoref{fig:resnet_unit}, where the left branch corresponds to the shortcut connection and the right branch is learning the residual mapping with a few convolution layers. Deep residual networks  are then built by stacking a large number of  these residual units.

While this residual learning architecture may seem like a trivial recasting of the mappings in the network, it has been found in practice to be much easier to train in deep networks \citep{He2015}, for several reasons. First and foremost, this architecture nearly eliminates the vanishing gradients problem by providing an unhindered route for the gradients to backpropagate through the layers using the shortcut connection. Another major advantage is that weight initialisation can be made more robust, given that the weights in a residual connection should be close to zero. Finally, the residual mappings are expected to be simpler than the full mappings,  leading to an easier overall optimisation problem.

In this work, we will adopt a specific type of units advocated by \cite{He2016}, the \emph{pre-activated bottleneck residual unit}. Pre-activation refers to the inversion  of the conventional ordering of convolution and activation functions. In a classical CNN architecture, the activation function is applied to the output of the convolution, as described in \autoref{eq:convnet}, but in a pre-activated architecture, the element-wise activation function is first applied to the input image, followed by the convolution. This alternative architecture is empirically found to yield better performances. These units are built around a stack of 1x1, 3x3, and 1x1 convolution layers (hence the `bottleneck' appellation).  This is  also empirically found to yield similar performance as a stack of two 3x3 convolutions but for a reduced number of parameters. \autoref{fig:residual_block} illustrates two variants of these residual units; \autoref{fig:resnet_unit} preserves  the  dimensions and depth of the input, while \autoref{fig:resnet_unit_dec} allows for a downsampling in resolution and an increase in depth by inserting a convolution layer in the shortcut branch. This downsampling of a factor 2 is performed  by using a strided convolution instead of using a pooling layer  as in a conventional CNN. Finally, these blocks use \emph{batch normalisation} layers, which, as their name implies, standardise their outputs by removing a mean value and  dividing by a standard deviation estimated during training. 

\section{CMU DeepLens}
\label{sec:deeplens}

Having provided some general background on Deep Learning in the previous section, we now present the details of our proposed \texttt{CMU DeepLens} strong-lens finder. In particular, we describe its architecture, training procedure and implementation.

\subsection{Architecture}

\begin{figure}
\centering
		\begin{tikzpicture}[  
	block/.style    = {draw, thick, rectangle, minimum height = 2em, minimum width = 4cm, node distance = 0.9cm},
    sum/.style      = {draw, circle, node distance = 1.cm}, % Adder
	input/.style    = {coordinate, node distance = 1.cm}, % Input
  	output/.style   = {coordinate, node distance = 1.cm} % Output
]
\definecolor{mycolor1}{HTML}{aadb32}
\definecolor{mycolor2}{HTML}{5bc862}
\definecolor{mycolor3}{HTML}{27ad80}
\definecolor{mycolor4}{HTML}{208f8c}
\definecolor{mycolor5}{HTML}{2c718e}
\definecolor{mycolor6}{HTML}{3b518a}
\definecolor{mycolor7}{HTML}{472b7a}
%Nodes
\node[input] (input1)  {Image};
\node[block, below of=input1, fill=mycolor1!30!white]  (layer1)   {Conv 7x7-32, ELU, B. N.};
\node[block, below = 0.5 of layer1, fill=mycolor2!30!white]  (layer2)   {ResNet-16-32};
\node[block, below of=layer2, fill=mycolor2!30!white]  (layer3)   {ResNet-16-32};
\node[block, below of=layer3, fill=mycolor2!30!white]  (layer4)   {ResNet-16-32};

\node[block, below = 0.5 cm of layer4, fill=mycolor3!30!white]  (layer5)   {ResNet-32-64, /2};
\node[block, below of=layer5, fill=mycolor3!30!white]  (layer6)   {ResNet-32-64};
\node[block, below of=layer6, fill=mycolor3!30!white]  (layer7)   {ResNet-32-64};.

\node[block, below = 0.5cm of layer7, fill=mycolor4!30!white]  (layer8)   {ResNet-64-128, /2};
\node[block, below of=layer8, fill=mycolor4!30!white]  (layer9)   {ResNet-64-128};
\node[block, below of=layer9, fill=mycolor4!30!white]  (layer10)   {ResNet-64-128};

\node[block, below =0.5cm of layer10, fill=mycolor5!30!white]  (layer11)   {ResNet-128-256, /2};
\node[block, below of=layer11, fill=mycolor5!30!white]  (layer12)   {ResNet-128-256};
\node[block, below of=layer12, fill=mycolor5!30!white]  (layer13)   {ResNet-128-256};

\node[block, below =0.5cm of  layer13, fill=mycolor6!30!white]  (layer14)   {ResNet-256-512, /2};
\node[block, below of=layer14, fill=mycolor6!30!white]  (layer15)   {ResNet-256-512};
\node[block, below of=layer15, fill=mycolor6!30!white]  (layer16)   {ResNet-256-512};

\node[block, below =0.5cm of layer16, fill=mycolor7!30!white] (layer17)  {fc 1, sigmoid};
 
%Lines
\draw[->, very thick] (input1) node [above] {Image} --  (layer1.north);
\draw[->, very thick] (layer1) -- (layer2);
\draw[->, very thick] (layer2) -- (layer3);
\draw[->, very thick] (layer3) -- (layer4);
\draw[->, very thick] (layer4) -- (layer5);
\draw[->, very thick] (layer5) -- (layer6);
\draw[->, very thick] (layer6) -- (layer7);
\draw[->, very thick] (layer7) -- (layer8);
\draw[->, very thick] (layer8) -- (layer9);
\draw[->, very thick] (layer9) -- (layer10);
\draw[->, very thick] (layer10) -- (layer11);
\draw[->, very thick] (layer11) -- (layer12);
\draw[->, very thick] (layer12) -- (layer13);
\draw[->, very thick] (layer13) -- (layer14);
\draw[->, very thick] (layer14) -- (layer15);
\draw[->, very thick] (layer15) -- (layer16);

\draw[->, very thick] (layer16) -- node [right] {avg pool} (layer17);

% Add comments

\node[right = 1cm of layer1,align=center] {output size:\\32x45x45};
\node[right = 1cm of layer2,align=center] {output size:\\32x45x45};
\node[right = 1cm of layer5,align=center] {output size:\\64x23x23};
\node[right = 1cm of layer8,align=center] {output size:\\128x12x12};
\node[right = 1cm of layer11,align=center] {output size:\\256x6x6};
\node[right = 1cm of layer14,align=center] {output size:\\512x3x3};
\node[right = 1cm of layer17,align=center] {output size :1};

\end{tikzpicture}
	\caption{Architecture of \texttt{CMU DeepLens}.  The  first block is a single convolution layer with an ELU activation function and batch normalisation, the following residual units  are illustrated on \autoref{fig:convnet}.The last block is a single fully connected layer with a sigmoid (logistic function) activation function, which outputs a probability between 0 and 1.}
	\label{fig:cmu_deeplens}
\end{figure}
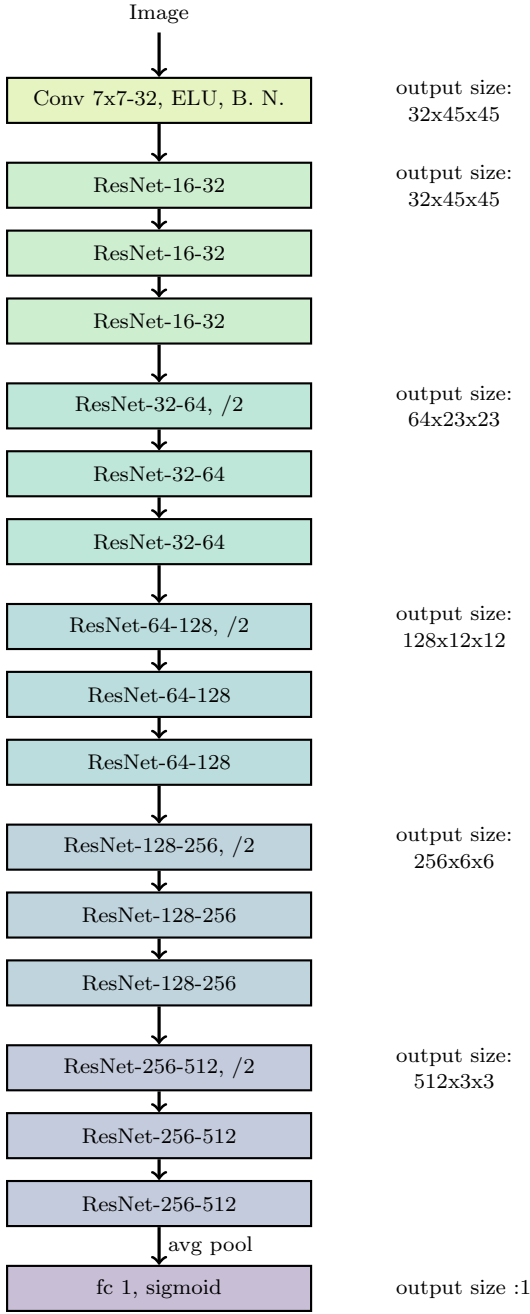

Our \texttt{DeepLens} model is a direct adaptation of the residual network architecture proposed in \citet{He2015, He2016}. The model described here corresponds to our fiducial choice of depth and complexity, which we found was suitable for the complexity of our task and the size of training data. 

The architecture of our model is illustrated in \autoref{fig:cmu_deeplens}. The input image is first processed by a convolution layer, which can accommodate single- or multi-band images, before going through a stack of pre-activated bottleneck residual units (illustrated in \autoref{fig:residual_block}), arranged in several levels of progressively coarser resolution.  In total, our model is 46 layers deep.

The output of the residual network is finally processed through a single fully-connected layer with a sigmoid activation function. Apart from the last layer, we use the Exponential Linear Unit (ELU) activation \citep{Clevert2015} throughout, which slightly differs from the ReLU activation:
\begin{equation}
	f(x ) = \begin{cases}
    x & \text{if } x\geq 0\\
    e^{x } - 1,              & \text{otherwise}
\end{cases}\;.
\end{equation}

If $y \in \{0,1\}$ is the class of an input image $\bm{x}$, the output of our model represents the estimated  probability $\hat{y} = q_{\bm{\theta}}( y = 1\  | \ \bm{x}) \in [0,1]$ of the input image $\bm{x}$ being a strongly lensed system (i.e. belonging to the class $y=1$), where $\bm{\theta}$ are the parameters of the  model (the weights of the neural network).  A detection will correspond to $\hat{y}$ crossing a given detection threshold, which will balance the trade-off between false and true detections as will be discussed in \autoref{sec:results}.

\subsection{Classification cost function}

For the binary classification problem of strong lens detection, we use the \emph{binary cross-entropy} cost function.  

Let $y\nnn{n}\in\{0,1\}$ denote the label of each instance $\bm{x}\nnn{n}$ and let $\hat{y}\nnn{n} \in [0, 1]$ be the probability
$q_{\bm{\theta}}(y\nnn{n}=1 \mid \bm{x}\nnn{n})$ estimated from the model .
The likelihood of the binary classification can be written as $\prod_n \hat{y}\nnn{n}^{y\nnn{n}}\times(1-\hat{y}\nnn{n})^{1-y\nnn{n}}$ which results in the following negative log-likelihood:
\begin{equation}
	-\sum_{n=1}^{N} y\nnn{n}\log \hat{y}\nnn{n} + (1-y\nnn{n})\log (1-\hat{y}\nnn{n}) \;,
   \label{eqn:cross_entropy}
\end{equation}
where $N$ is the number of training instances. This loss function can also be interpreted  as the cross entropy between $p(y \mid \bm{x})$ and our model $q_{\theta}(y \mid \bm{x})$. The network is then trained to minimise this cross entropy (negative log likelihood) objective.  

\subsection{Preprocessing and data augmentation}
\label{sec:preprocess}

We apply minimal pre-processing to the input images. We subtract the mean image from the training set and normalise the images by the noise standard deviation $\sigma$ in each band evaluated over the whole dataset. Finally, extreme values are clipped to restrict the dynamic range of the input images to within a given $k \sigma$ (we use here $k=250$, but this value can be adjusted).  Although these steps are usual and sensible pre-processing techniques, we do not find a significant impact on the results if they are omitted. If masked pixels are present in the image, they are set to 0 as a last stage of preprocessing (i.e. after  mean subtraction and normalisation).

More crucial to effectively train the  model, we further perform several data augmentation steps, as a way to increase the effective size of the training set and to make the model invariant to specific transformations. In particular, we want our model to be rotationally invariant, for which we apply random rotations to the training images. The  following steps are applied to the training data:
\begin{itemize}
	\item Random rotation of the image in the range $[-90, \ 90^\circ]$, using a spline interpolation scheme.
	\item Random mirroring along the vertical and horizontal axes.
	\item Random zooming of the image in the range $[0.9, 1]$, meaning that the image is randomly compressed (or stretched) by a factor within this range.
\end{itemize}
When these operations access pixels outside of the input image a simple wrapping strategy is used, and the augmented image remains the size of the original image. 

\subsection{Training procedure}
\label{sec:training}

We initialise the weights of our model with random normal values using the strategy proposed in \citet{He2015a} and all layers are trained from scratch. The network is trained over 120 epochs (i.e. passes over the whole training set) in mini-batches of 128 images using  ADAM \citep{Kingma2015} using the default exponential decay rates and a staring learning rate of $\alpha=0.001$, which is divided by $10$ every 40 epochs.

\subsection{Implementation}

Our model itself is implemented using the \texttt{Theano}\footnote{\url{http://deeplearning.net/software/theano/}} and \texttt{Lasagne}\footnote{\url{https://github.com/Lasagne/Lasagne}} libraries. We make our code publicly available at:
\url{https://github.com/McWilliamsCenter}

The results presented in this work were obtained on an Nvidia Titan X (Pascal) GPU. As will be described in the next section, we trained our model on a set 16,000 images of size $45 \times 45$ pixels. Despite the relatively small size of our training set, we find our data augmentation scheme to be very effective and we do not find any evidence of overfitting. 
On this dataset and hardware, the training procedure requires approximately 1 hour. Once the network is trained however, the classification itself is extremely efficient, requiring approximately $350 \mu s$ per image.

\section{Strong Lensing Simulations}
\label{sec:sims}
\begin{figure*}
\begin{subfigure}[t]{0.5\textwidth}
\includegraphics[width=\textwidth]{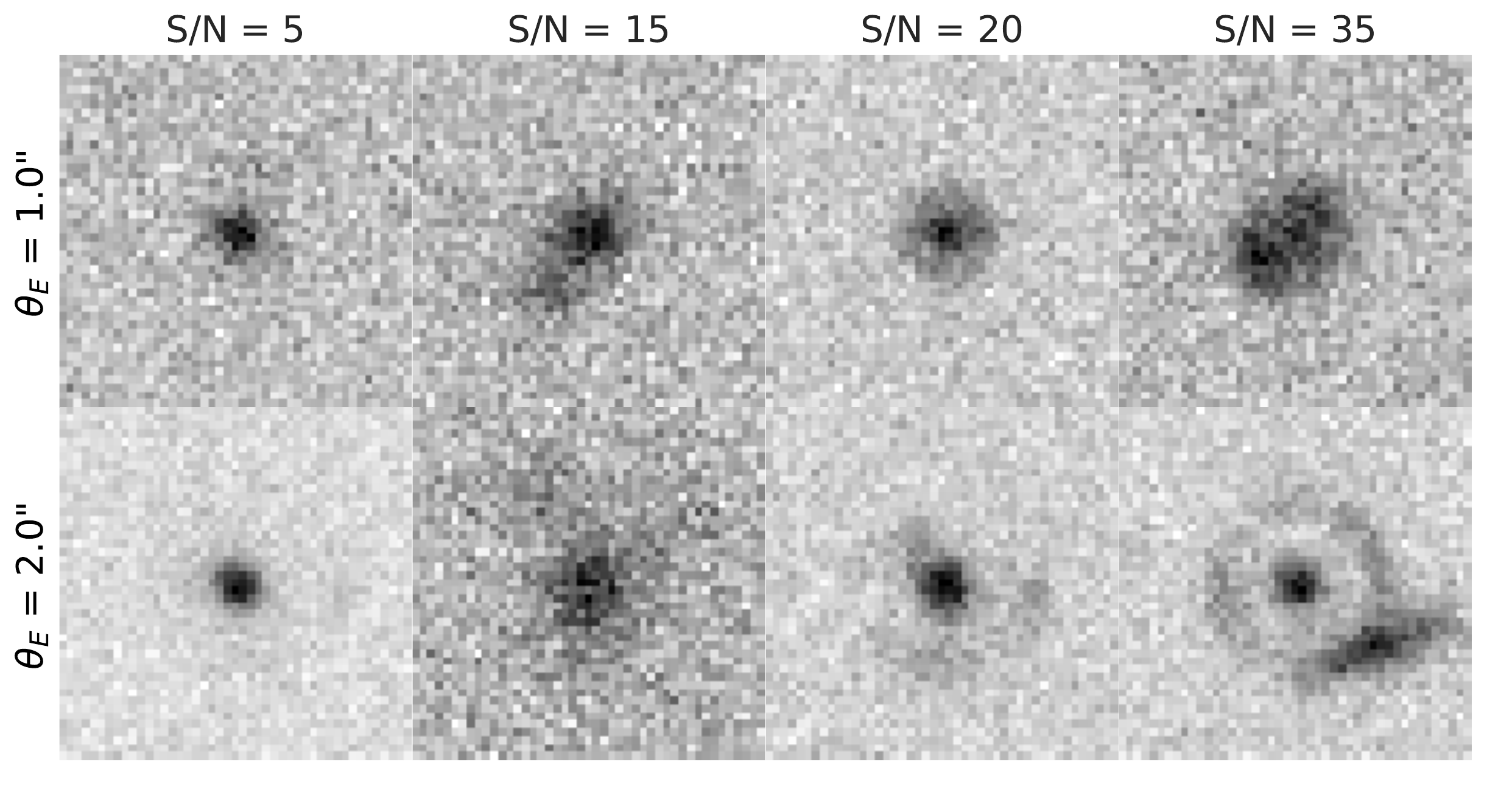}
\caption{Best single epoch}
\end{subfigure}~
\begin{subfigure}[t]{0.5\textwidth}
\includegraphics[width=\textwidth]{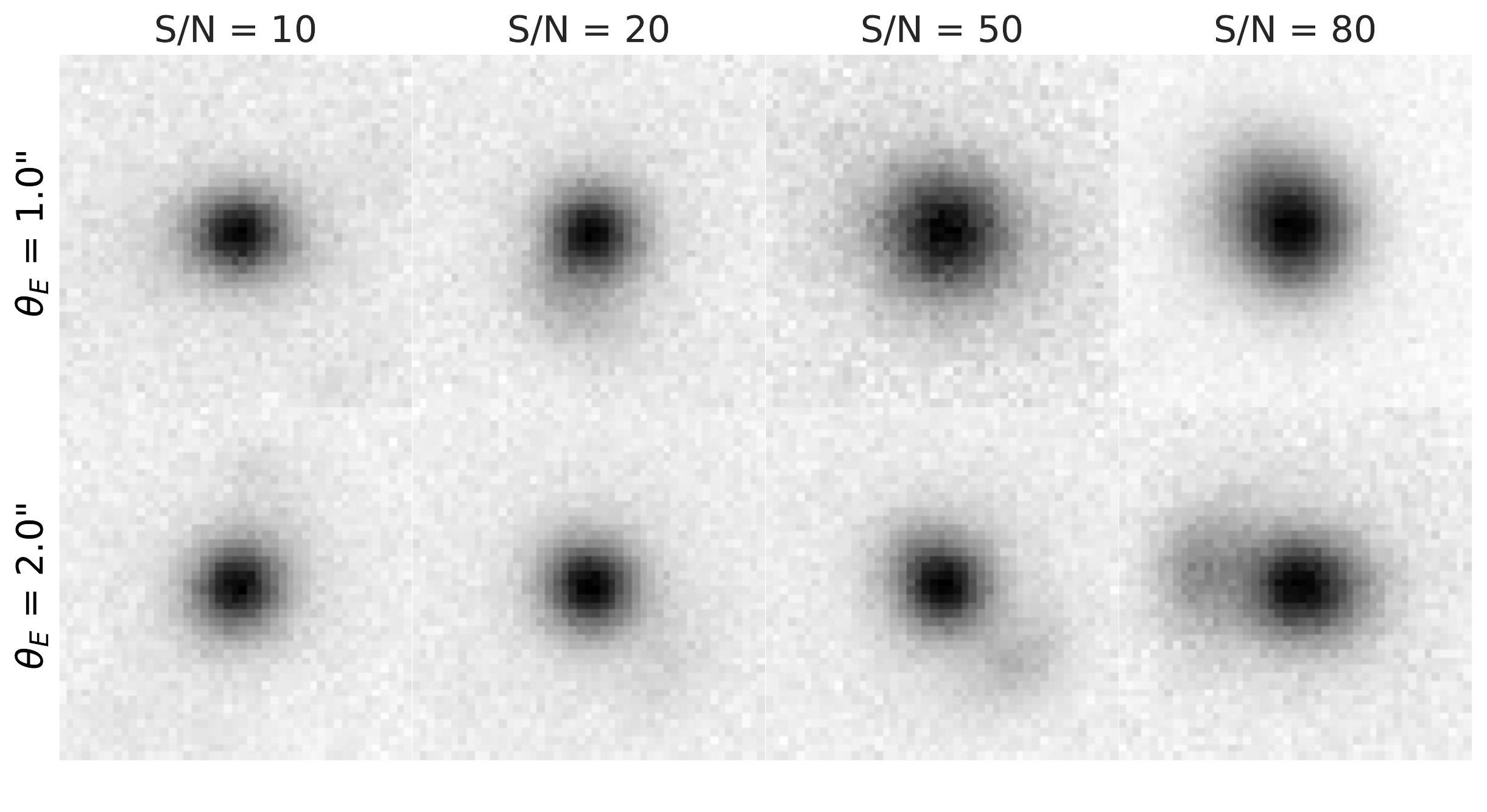}
\caption{Stack images}
\end{subfigure}
\caption{Randomly selected simulated lens images of various Einstein radius $\theta_E$ and arc S/N levels. Single-epoch images (left) are produced by using only the best seeing exposure while the stack images are produced by co-adding all exposures down to the worst seeing in the stack.}
\label{fig:lens_samples}
\end{figure*}

In this section, we detail the simulation pipeline that was used to produce training and testing sets for our lens finding model. We simulate LSST images of strong lenses using PICS (Pipeline for Images of Cosmological Strong lensing; \citealt{Li2016}) for the strong lensing ray-tracing and LensPop\footnote{\url{https://github.com/tcollett/LensPop}} \citep{Collett2015} for mock LSST observing.

All simulated images contain a central early-type galaxy as well as some additional galaxies in the field of view. In half of the simulated images, we further include a strongly lensed background galaxy. 

\subsection{Lens galaxy model}

For these simulations, we use a population of elliptical lens galaxies, as they are expected to dominate the population of galaxy-scale strong lenses \citep{Oguri2010}. To model the mass profile of these elliptical galaxies, we assume a Singular Isothermal Ellipsoid (SIE) profile. This model is not only analytically tractable, it has also been found to be consistent with models  of individual lenses and lens statistics on the length scales relevant for strong lensing \citep[e.g.][]{Koopmans2006, Gavazzi2007,Dye2008}. The SIE density profile is defined as:
\begin{equation}
\rho(x, y) = \frac{\sigma_v^2}{2 \pi G \left( x^2/q + q y^2 \right)} \;,
\end{equation}
where $\sigma_v$ is the velocity dispersion of the lens and $q$ is the axis ratio of the ellipsoid. For such a profile, the convergence of the lens is given by:
\begin{equation}
	\kappa(x, y )= \frac{\theta_{\rm E}}{2}\frac{1}{\sqrt{x^2/q+y^2 q}} \;,
\end{equation}
where $\theta_{\rm E}$ is the Einstein Radius which can be calculated according to the redshift of the lens $z_l$, the redshift of the source $z_s$, and the velocity dispersion of the lens galaxy as:
\begin{equation}
\theta_{\rm E} = 4\pi\left(\frac{\sigma_v}{c}\right)^2 \frac{D_{ls}(z_l,z_s)}{D_{s}(z_s)} \;, 
\end{equation}
where $c$ is the speed of light, $D_{ls}$ and $D_{s}$ are the angular diameter distance from the source plane to the lens plane and from the source plane to the observer respectively. 

Given these relations, a lens is entirely described by only a few parameters: $\{\sigma_v, q, z_l, z_s, \phi \}$, where $\phi$ is a rotation angle. We uniformly sample the velocity dispersion, ellipticity and orientation of the lenses from a range of typical values, so that $\sigma_v \in [150, 350 \ \text{km/s}] $, $q \in [0.5, 1.0]$, and $\phi = [0, 2\pi]$. The lens redshifts are obtained by matching the velocity dispersion in our simulations to the catalog of elliptical galaxies in the COSMOS survey from \cite{Zahid2015}. The resulting redshift range of our lenses is $z_l \in [0.2, 0.7]$. The source galaxies will be assumed to be at a fixed redshift of $z_s = 2.0$.

To model the light distribution of the lens galaxies, we use an elliptical S\'ersic profile, defined as:
\begin{equation}
I(R) =  I_{\rm eff}~{\rm exp} \left\lbrace -b_{n} \left[ \left( \frac{R}{R_{\rm eff}}\right)^{1/n} - 1 \right ] \right\rbrace \;,
\end{equation}
where $R = \sqrt[]{x^2 /q+y^2 q }$, $R_{\rm eff}$ is the effective radius in units of arcsecond, $I_{\rm eff}$ is the intensity at the effective radius, $n$ is the index of the S\'ersic profile. For each lens galaxy, we use for the light profile the same axis ratio and orientation as the SIE mass profile and we set the effective radius, luminosity and S\'ersic index to the matched COSMOS galaxy.

\subsection{Background sources and additional line-of-sight galaxies}

We use for the lensed background sources a set of detailed images of low-redshift bright galaxies ($z \sim 0.45$) extracted from  the mosaics produced by the  CANDELS team \citep{Grogin2011, Koekemoer2011} and selected from the CANDELS UDS catalog \cite{Galametz2013}. These galaxies are rescaled and placed at a fixed redshift $z_s = 2.0$, near the caustics of the lensing system so as to produce lensed arcs. The lensed images themselves are then produced by ray-tracing simulations as part of the PICS pipeline. 

To add to the complexity of the generated image and make them look more realistic, we further populate the field with galaxies drawn from the Hubble Ultra Deep Field. These galaxies are placed along the line of sight but in these simulations, their images are not lensed even if they are located behind the lens.

Note that our simulated fields contain a single deflector on a single lens plane, we do not include any additional perturbative weak lensing nor any compound lensing effects \citep[e.g.][]{Collett2016}.

\subsection{Mock LSST observations}
\begin{table}
	\centering
	\caption{Parameters of the mock LSST observations. Only the median values are reported for the seeing and sky-brightness.}
	\label{tab:simulation}
	\begin{tabular}{ccccc}
		\hline
		\hline
		Band & Seeing & Sky & Exposure Time & Pixel Scale\\
		 & (arcsec) &     & (s) & (arcsec) \\
		\hline
		 $g$ & 0.81 & 21.7 & 3000 & 0.18 \\
		\hline
	\end{tabular}
\end{table}

To produce LSST-like images from the ideal images produced in the previous steps, we use the LensPop software \cite{Collett2015} to perform mock observing. Images are re-sampled to match the detector pixel scale and convolved with a circularly symmetric Gaussian Point Spread Function discretised at the same pixel scale.

A noisy realisation of this image is generated assuming a Poisson model based on the sky plus signal, and an additional Gaussian read-out noise. Parameters for these simulations follow \cite{Collett2015} and are based on the LSST observation simulator \citep{Connolly2010}. They are summarised in \autoref{tab:simulation}.

To account for seeing and sky-brightness variations over the course of the survey, each simulated exposure is drawn from a stochastic distribution of these parameters. We then consider two different strategies to use these exposures. For each field, we build one \emph{single-epoch} image by keeping only the best seeing exposure and another ``worst-case'' \emph{stack} image by co-adding all individual exposures after degrading them all in resolution to match the one with the worst seeing. These two sets of images will allow us to investigate the trade-off between signal-to-noise and resolution for our automated lens search. 

Examples of final mock observations with these two strategies are shown on \autoref{fig:lens_samples} for different arc sizes and signal to noise ratios (see  \autoref{eq:snr} for our definition of S/N). This ratio is  defined in our final images in terms of the total flux of the lensed arc:
\begin{equation}
	S/N = \frac{F_{tot}}{\sigma \sqrt{N_{pix}}}
	\label{eq:snr}
\end{equation}
where $F_{tot}$ is the total flux of the lensed source, $\sigma$ is the rms noise in the observed image and $N_{pix}$ is the number of pixels associated with the lensed source (measured by segmentation on a noiseless image of the arc using a threshold of $0.5 \ \sigma$, where $\sigma$ is the rms noise in the corresponding full image.).

\section{Results}
\label{sec:results}
\begin{figure*}
		\begin{subfigure}[t]{\textwidth}
		\includegraphics[width=0.32\textwidth]{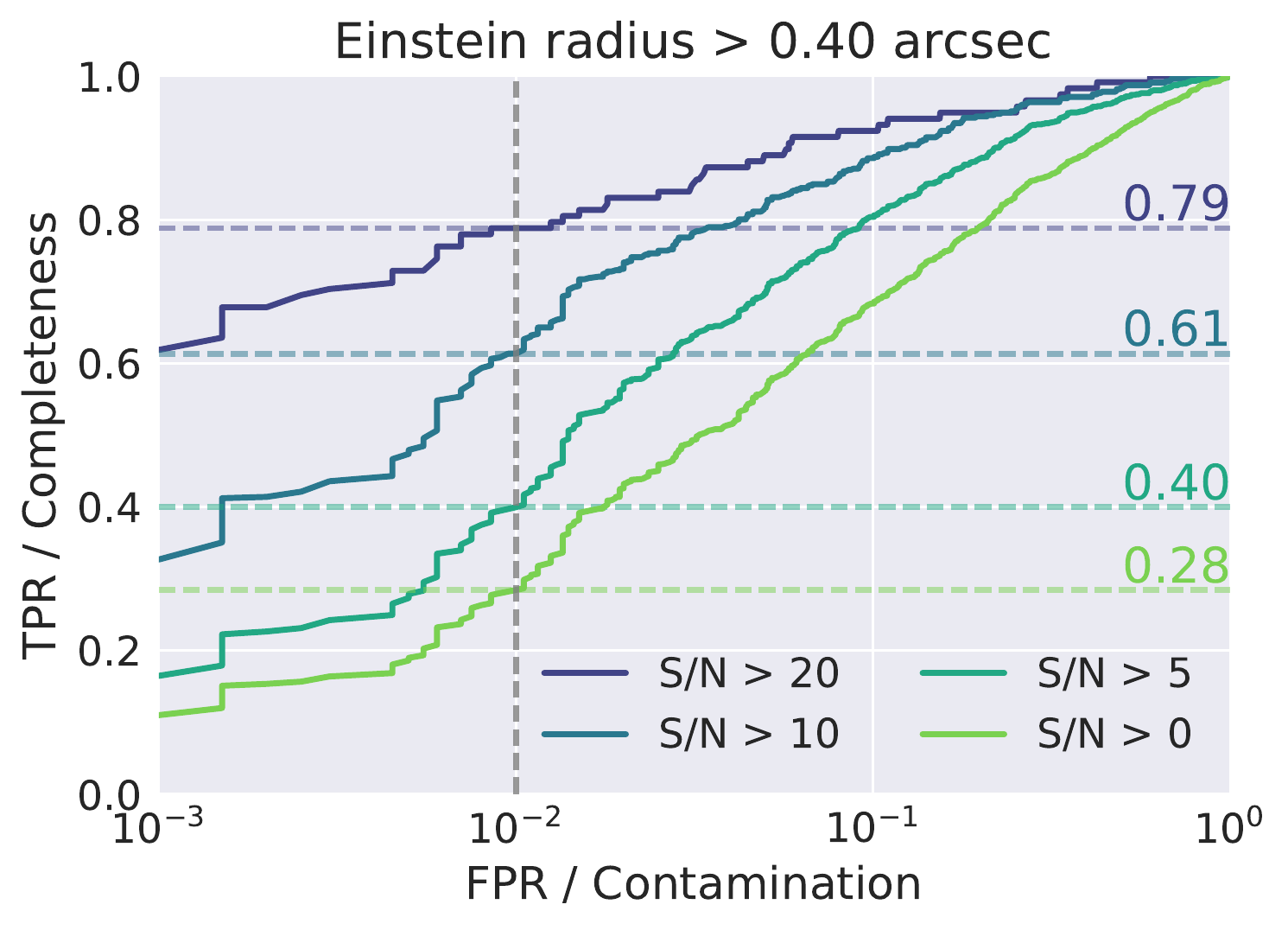}~
		\includegraphics[width=0.32\textwidth]{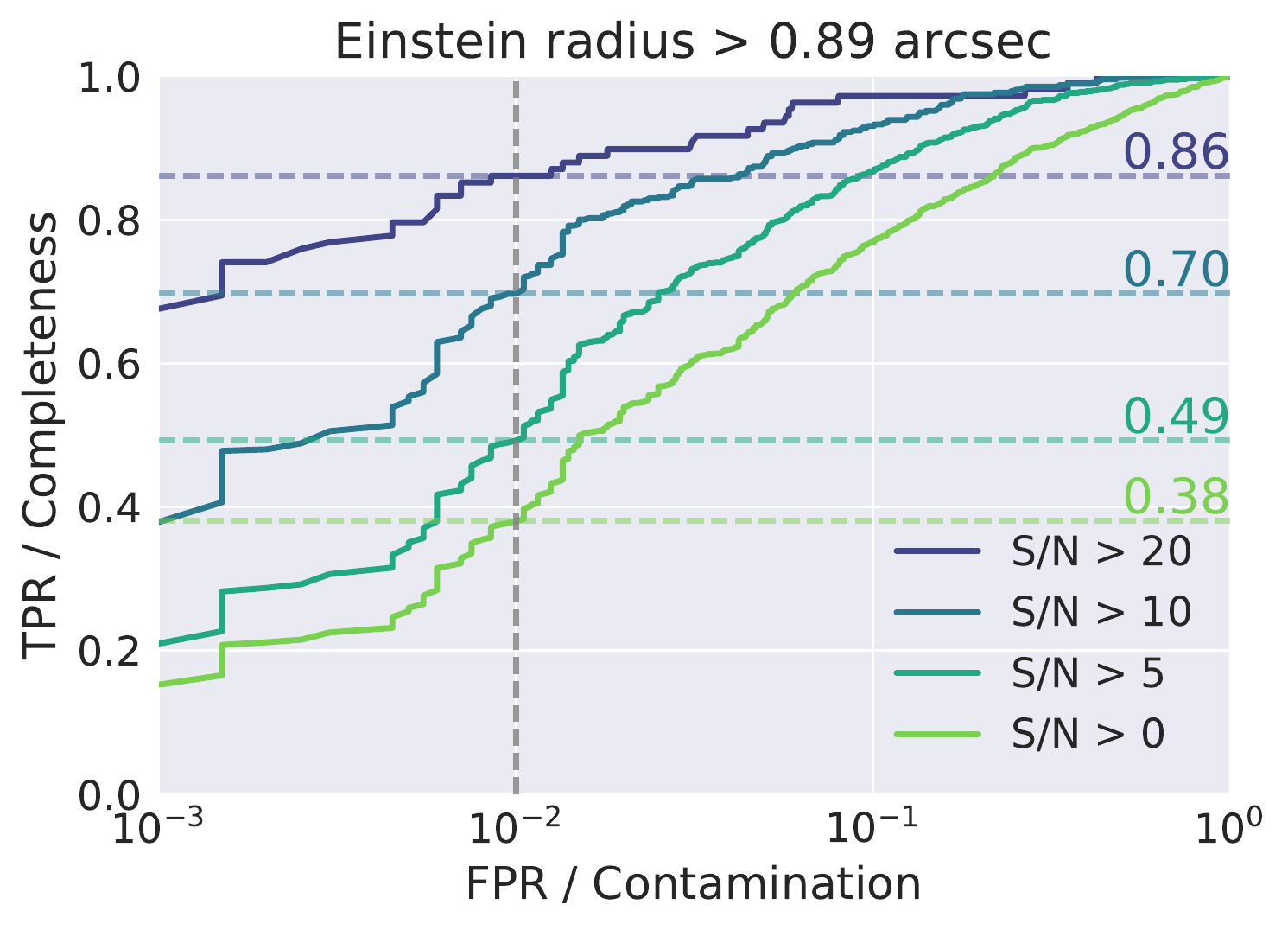}~
		\includegraphics[width=0.32\textwidth]{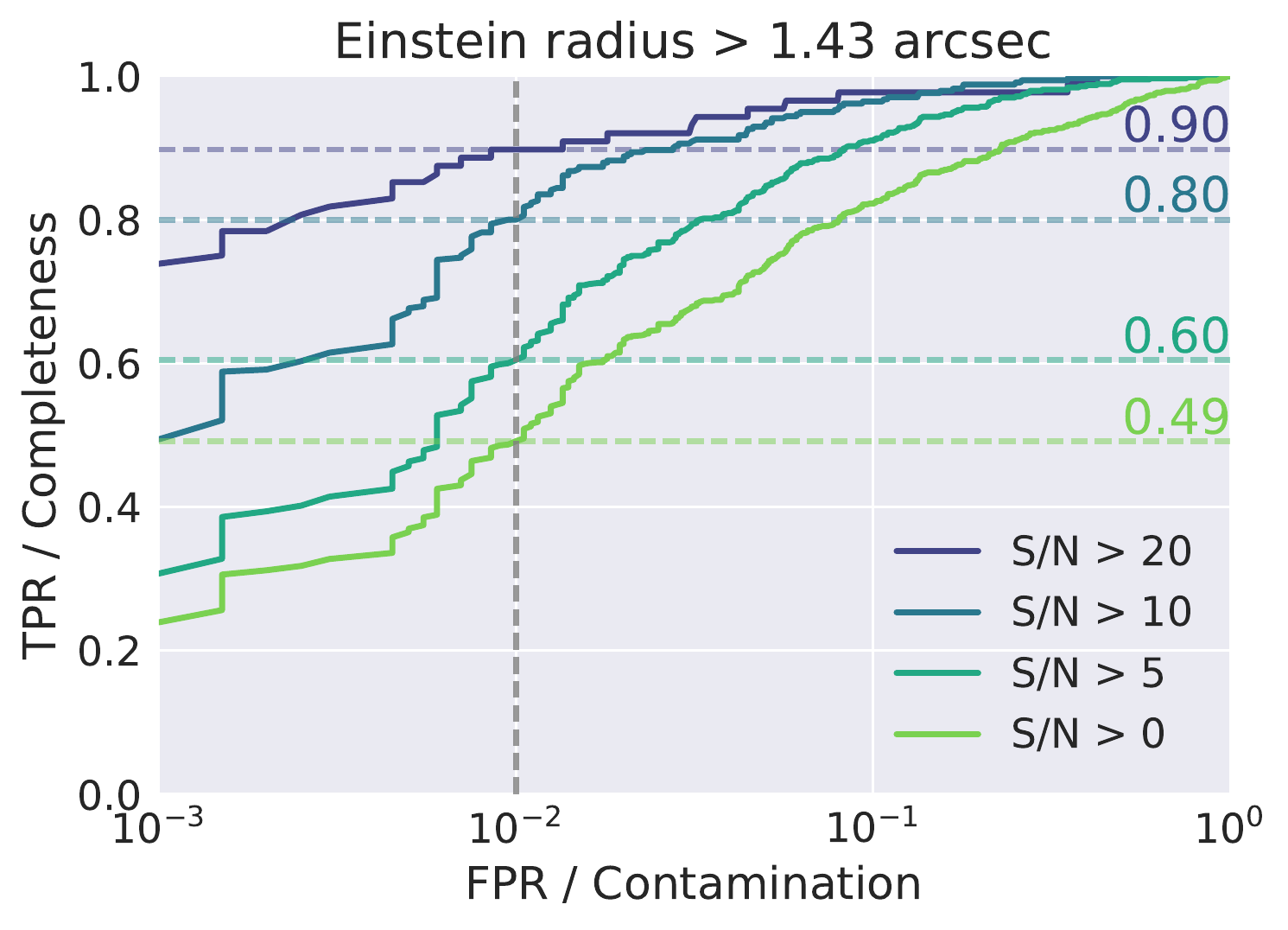}
		\caption{Single best epoch images}
	\end{subfigure}\\
	\begin{subfigure}[t]{\textwidth}
		\includegraphics[width=0.32\textwidth]{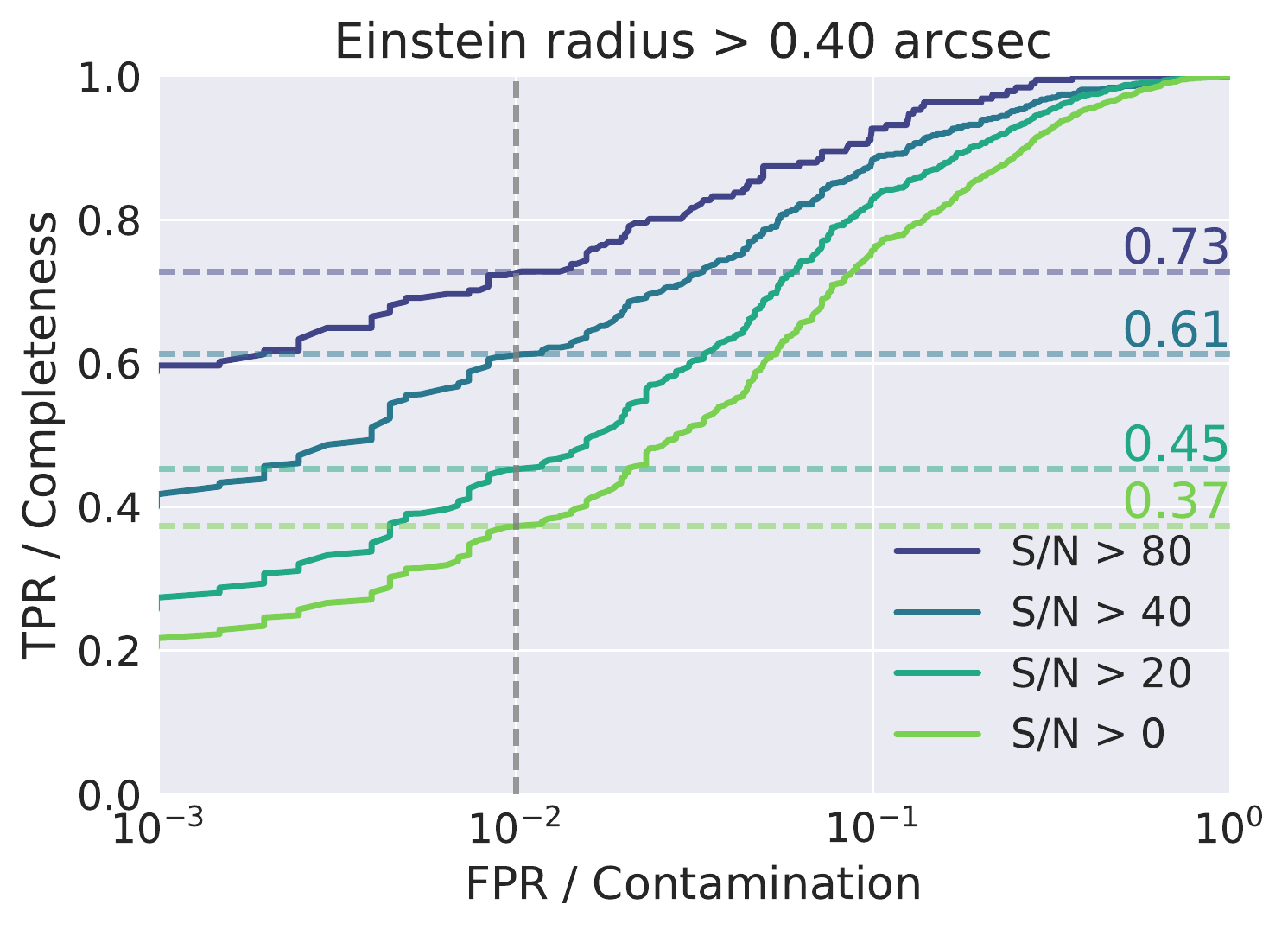}~
		\includegraphics[width=0.32\textwidth]{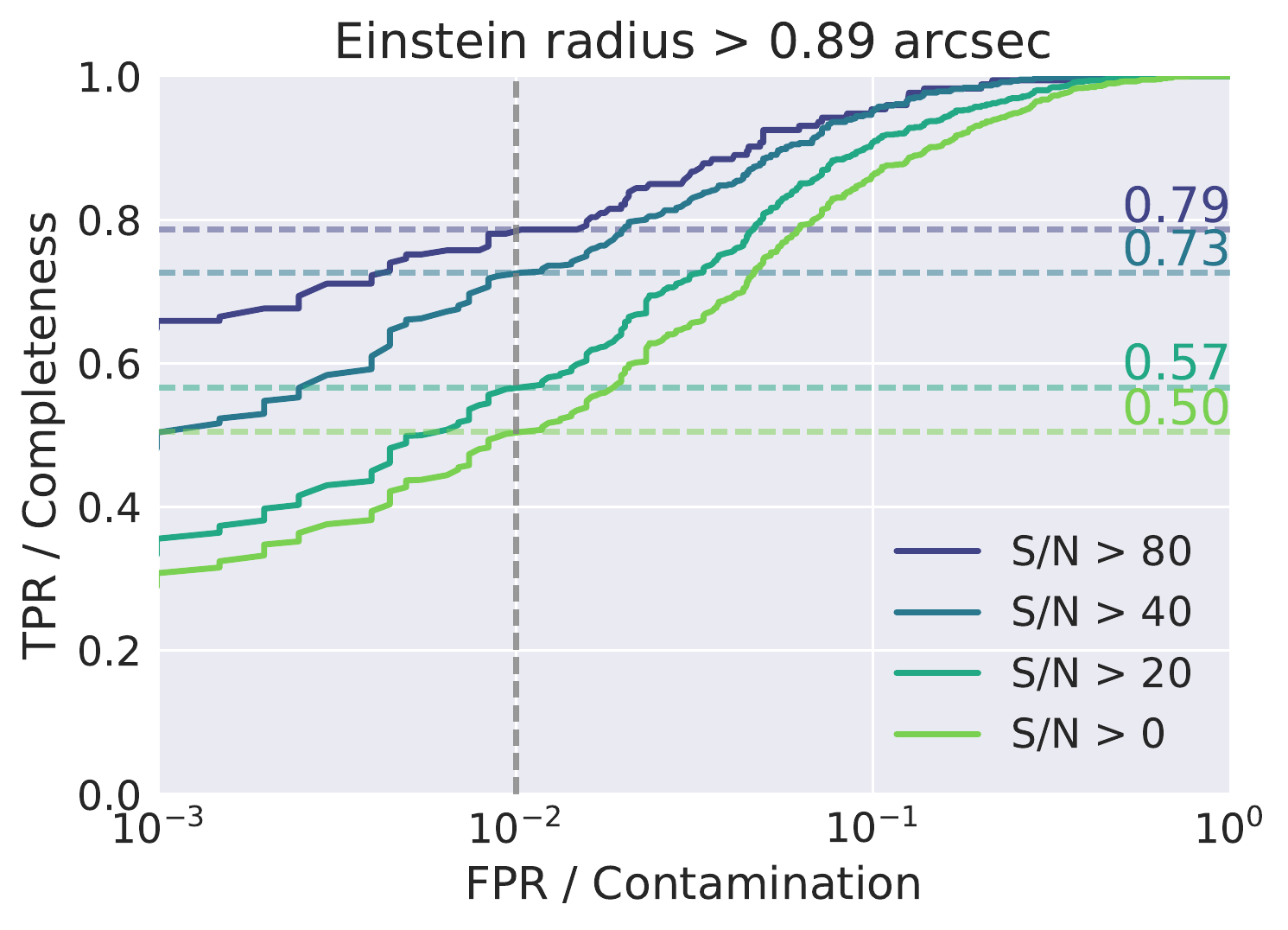}~
		\includegraphics[width=0.32\textwidth]{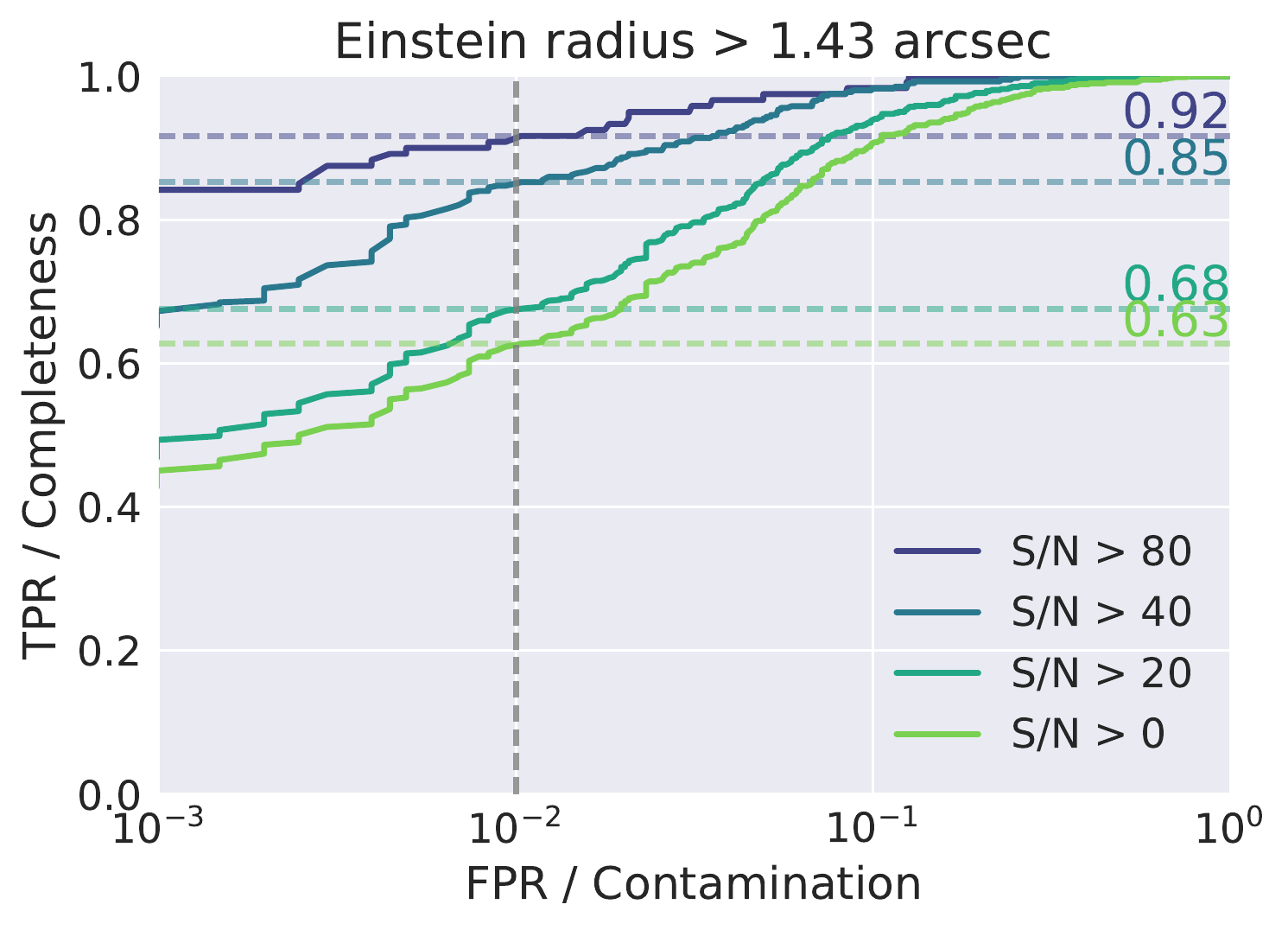}
		\caption{Stack images}
	\end{subfigure}
	\caption{ROC curves for different cuts in Einstein radius and S/N of the input lensed image. The dashed vertical lines correspond to our fiducial 1\% contamination threshold while the horizontal dashed lines indicate the corresponding completeness for that contamination threshold. As we exclude from the input sample fainter and smaller lenses, the completeness progressively increases.}

	\label{fig:purVcomp}
\end{figure*}

\begin{figure}
	\begin{subfigure}[t]{\columnwidth}
	\includegraphics[width=0.5\columnwidth]{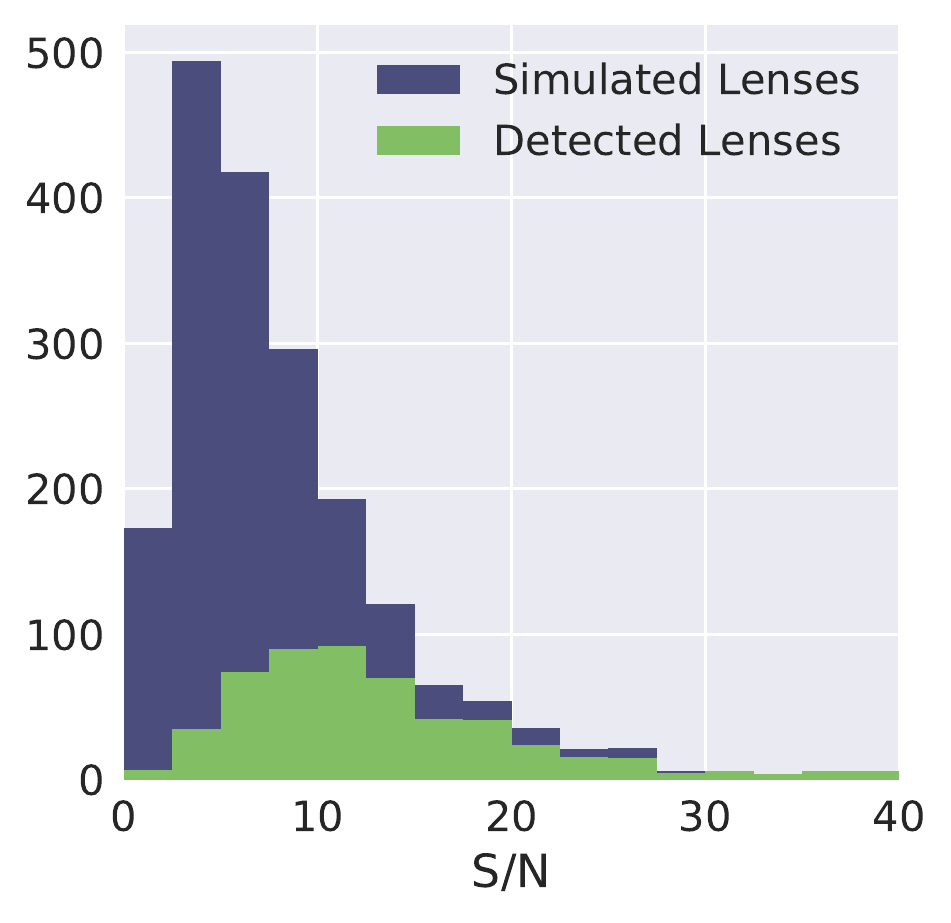}~
	\includegraphics[width=0.5\columnwidth]{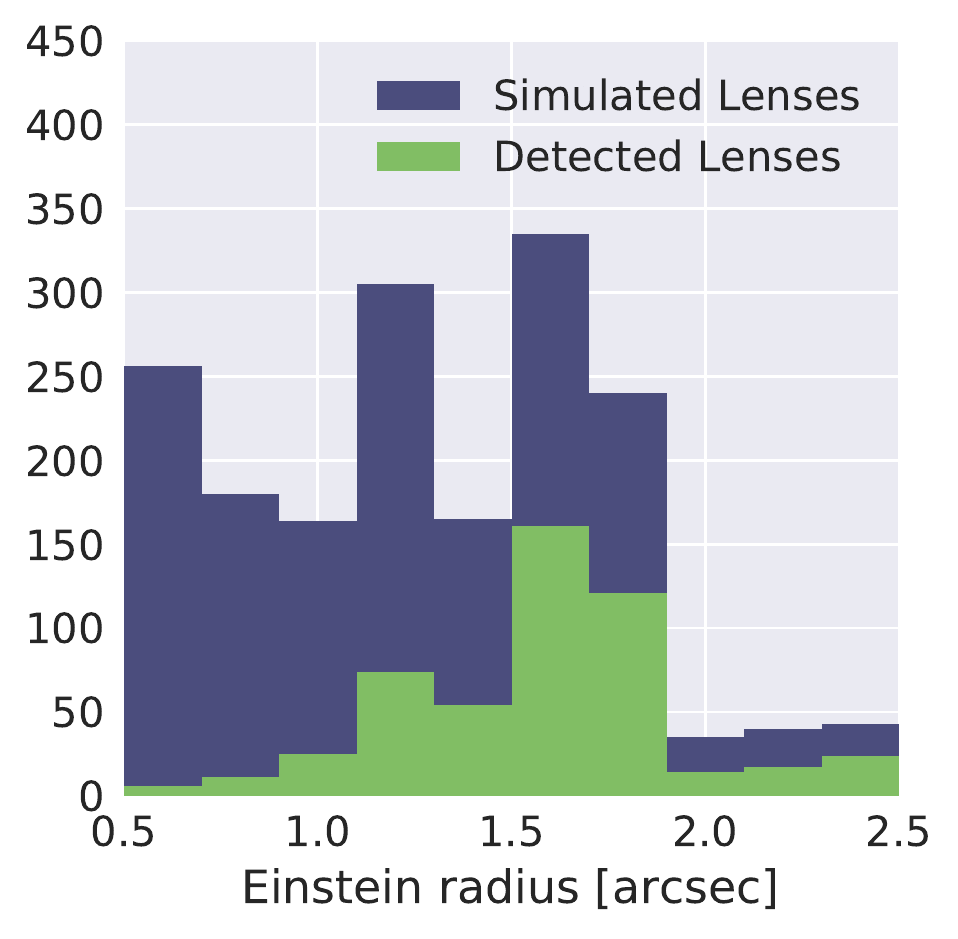}
    \caption{Single best epoch images}
	\end{subfigure}\\
	\begin{subfigure}[t]{\columnwidth}
	\includegraphics[width=0.5\columnwidth]{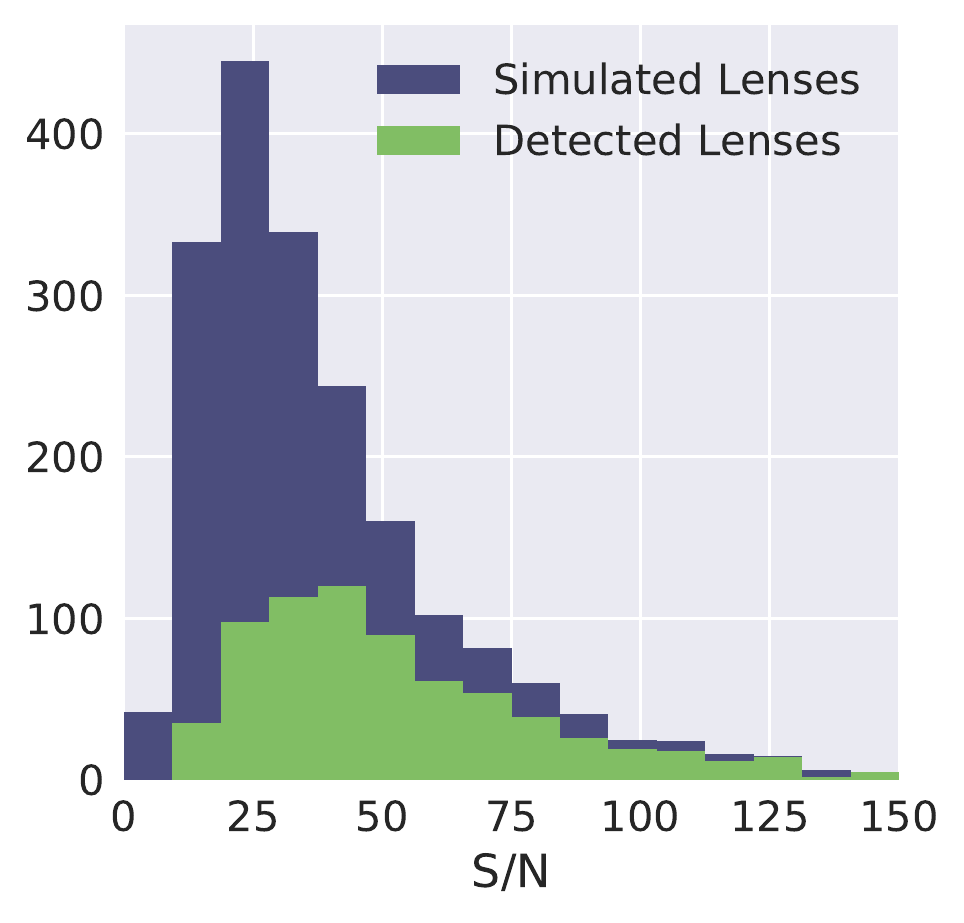}~
	\includegraphics[width=0.5\columnwidth]{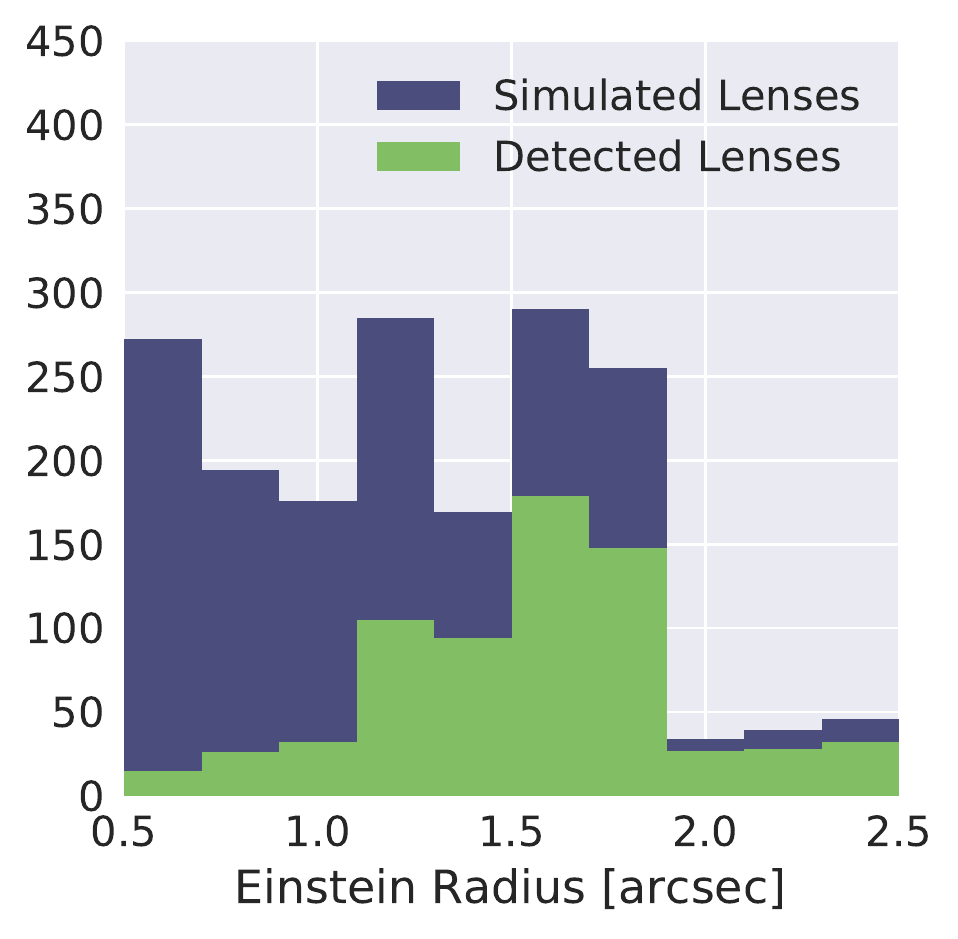}
    \caption{Stack images}
	\end{subfigure}
	\caption{S/N and Einstein radius distribution of the simulated and recovered lens populations, for our fiducial 1\% FPR detection threshold.}
    \label{fig:lens_recovery}
\end{figure}

In this section, we test the performance of our lens finder on the simulations described in the previous section. We perform these tests as a function of  Einstein radius $\theta_E$ and S/N of the lensed source. For each class of simulations (single-epoch and stack), we train a model on a random subset of 16,000 images, following the procedure described in \autoref{sec:training}. We then evaluate the performance of the classifier on a test set of 4,000 images as a function of various cuts in Einstein radius and S/N. 

To quantify  the performance of our lens finder, we measure the True Positive Rate (TPR) and False Positive Rate (FPR).  The TPR, also known as \emph{completeness} or recall, is defined as the ratio of detected lenses to the total number of lenses:
\begin{equation}
    \mbox{TPR} = \frac{N_{\mbox{true positives}}}{N_{\mbox{true positives}} + N_{\mbox{false negatives}}} \;.
\end{equation}
The FPR, which can also be interpreted as a \emph{contamination} rate is defined as the fraction of non-lens images wrongly identified as lenses:
\begin{equation}
	\mbox{FPR} = \frac{N_{\mbox{false positives}}}{N_{\mbox{false positives}} + N_{\mbox{true negatives}}} \;.
\end{equation}
This statistic gives us a handle on the expected contamination of a lens sample while being independent of the ratio of lens to non-lens images in the testing set (not representative of a real survey in our simulations). 

These statistics are a function of the detection threshold applied to the probability of an image containing a lens outputted by the model. This threshold balances completeness versus contamination of the lens sample. This trade-off is typically illustrated by the Receiver Operating Characteristic (ROC) curve which is obtained by plotting the TPR as a function of the FPR while varying this detection threshold from 0 to 1. \autoref{fig:purVcomp} shows the ROC curves of our classifier evaluated on the testing set, for various cuts in S/N and Einstein radius in the lens sample. We use a fiducial FPR value of 1\% to derive a detection threshold for the rest of this analysis. This FPR value, illustrated by a vertical dashed line on \autoref{fig:purVcomp}, would be set in practice based on what would be considered to be an admissible level of contamination in a given survey. The detection threshold corresponding to this FPR would then be derived from simulations. 

For this fiducial detection level, we compare the corresponding completeness achieved by \texttt{DeepLens} for samples of increasingly larger and brighter arcs. These different completeness levels are marked by horizontal dashed lines on \autoref{fig:purVcomp}.

We find that our lens finder exhibits very similar behaviour for the two sets of images. The increase in signal to noise seems to roughly compensate the loss in resolution. We note that in single exposure images, we achieve a completeness of 90\% for our fiducial choice of 1\% FPR, when considering arcs larger than $1.4$\arcsec. We reach similar completeness levels in stacked images for arcs of that size with S/N larger than 80.

To further illustrate the impact of S/N and resolution on the recovery of lenses we show in \autoref{fig:lens_recovery} the distributions of S/N and Einstein radius of our simulated lens population as well as the distribution of these properties in the recovered sample for our fiducial 1\% FPR threshold. Interestingly, we find that in stacked images, \texttt{DeepLens} can still recover some poorly resolved lenses with Einstein Radius lower than the median seeing.

\begin{figure*}
		\begin{subfigure}[t]{0.5\textwidth}
			\includegraphics[width=\textwidth]{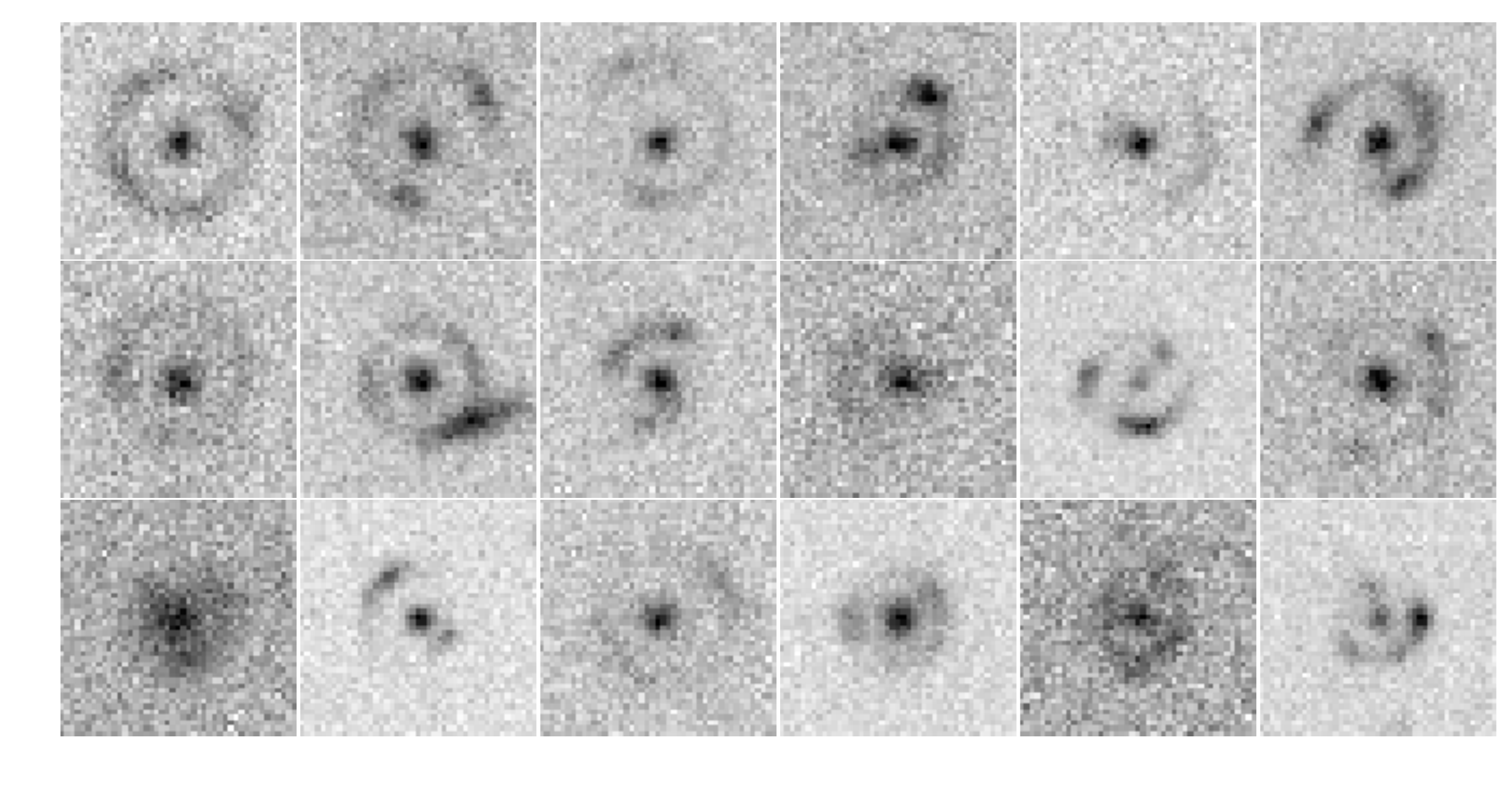}
			\caption{True positives single images}
		\end{subfigure}~
		\begin{subfigure}[t]{0.5\textwidth}
		\includegraphics[width=\textwidth]{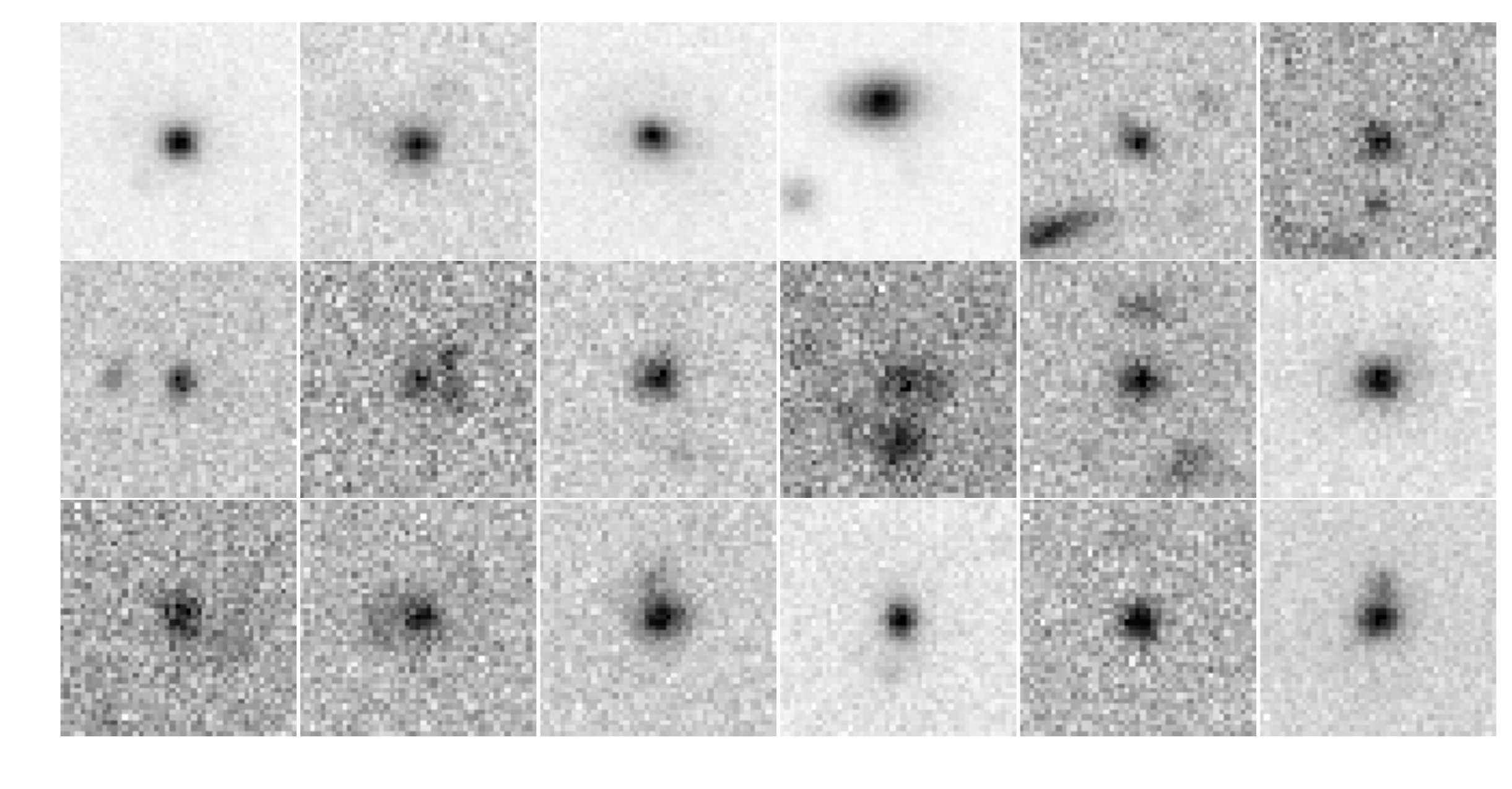}
			\caption{False positives single images}
		\end{subfigure}\\
		\begin{subfigure}[t]{0.5\textwidth}
		\includegraphics[width=\textwidth]{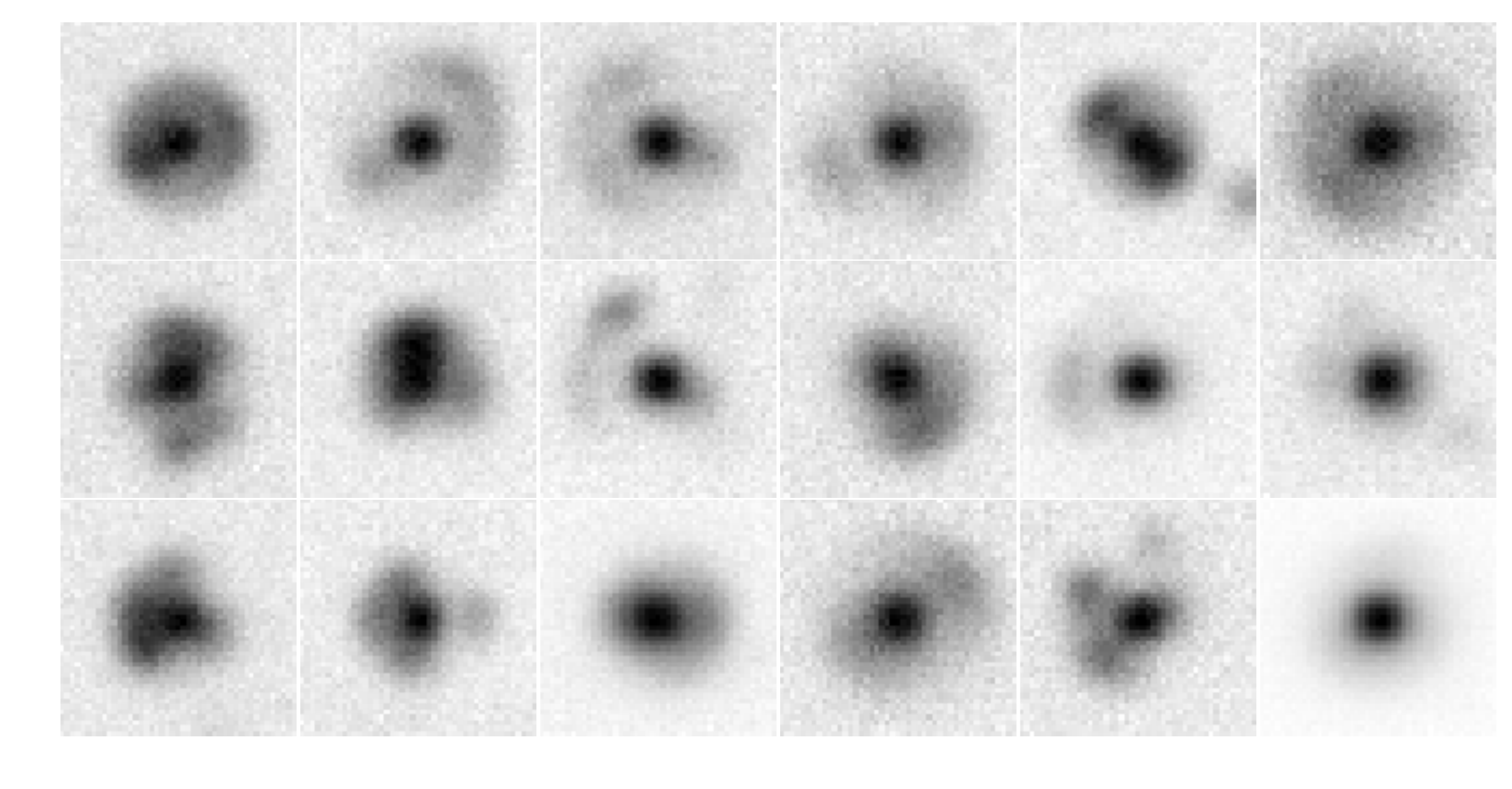}
		\caption{True positives stack images}
		\end{subfigure}~
		\begin{subfigure}[t]{0.5\textwidth}
		\includegraphics[width=\textwidth]{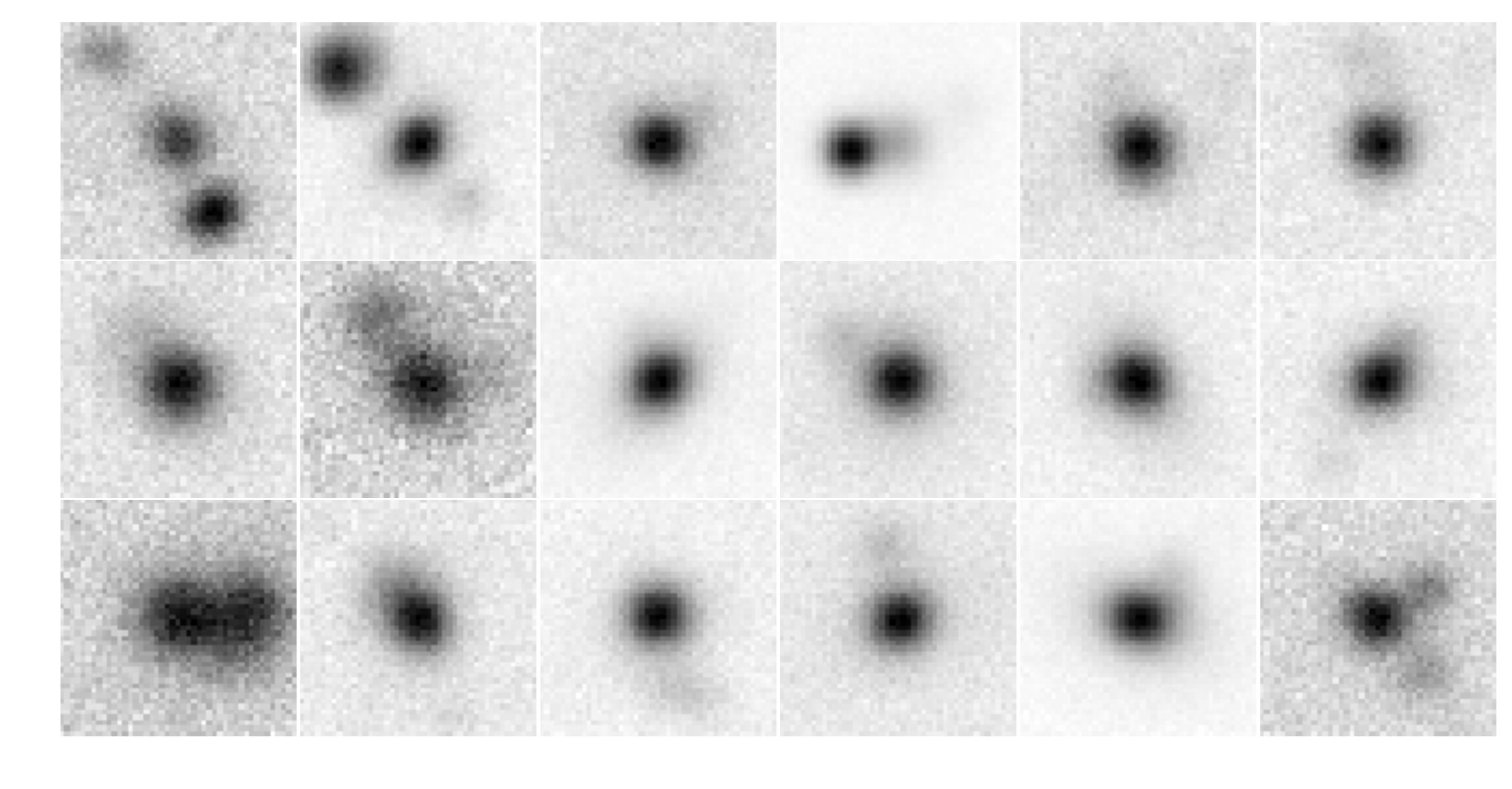}
		\caption{False positives stack images}
		\end{subfigure}
		\caption{Illustrations of true and false positives with the highest predicted lens probability in our two sets of simulations.}
		\label{fig:failure}
\end{figure*}
To visually investigate some failure modes of the model, we show on \autoref{fig:failure} some examples of true and false positives for our two sets of simulations. We retain on this figure only the images with the highest predicted lens probability. As can be seen, the lenses for which \texttt{DeepLens} is the most certain have clearly visible rings and multiple images. False positives are dominated by the presence of multiple objects at the vicinity of the lens. We expect these types of failures to be dramatically reduced in multi-band images, as the colour would provide the necessary additional information to discriminate real lenses from these false positives.

\section{Discussion}
\label{sec:discussion}

In this section we discuss the importance of simulations in a supervised machine learning method and provide several avenues to further improve our proposed model.

As in any supervised machine learning approach, the quality of the training set is a major factor in the performance of the method on actual data. For instance, \cite{Petrillo2017} find with a conventional CNN approach that most of the false positives that the method produces on real KiDS images come from contaminants such as ring galaxies, mergers or star-forming rings. For practical applications it is paramount to train Deep Learning models on representative datasets including all the diversity and variability of real survey images. We should therefore stress that the reported completeness and contamination levels on our simulations are only optimistic estimates of the performance of our model on real data.

\begin{figure}
	\includegraphics[width=\columnwidth]{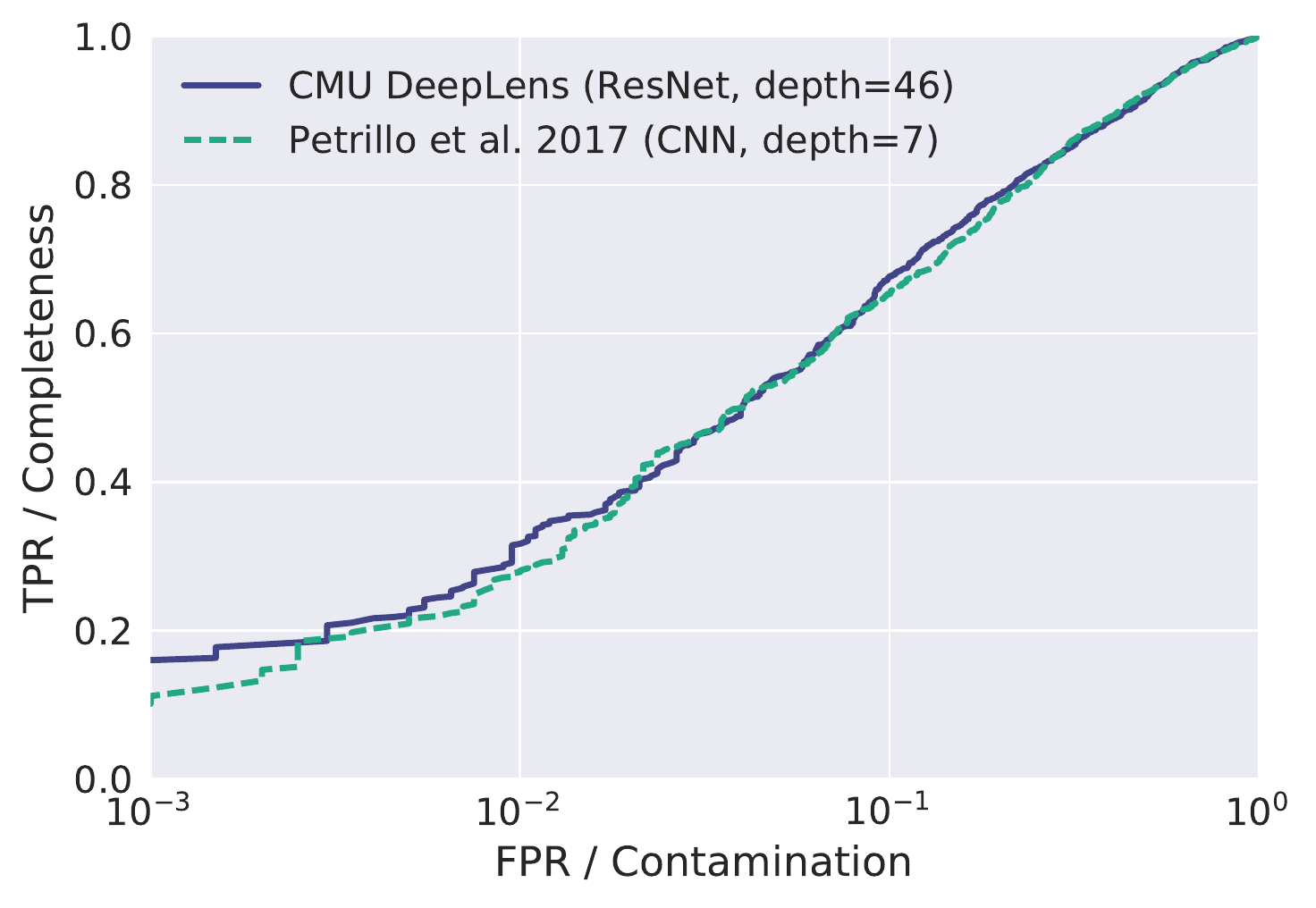}
    \caption{Comparison of ROC curves between our proposed model and the CNN proposed in \citet{Petrillo2017}, evaluated on the same training and testing sets of single epoch images. Note that these curves are computed on our full lens sample, without S/N and Einstein radius cuts.}
    \label{fig:resnet_vs_cnn}
\end{figure}

Realistic simulations are not only important for the final application to real data but also for method development. In particular we find that our simulations are not complex enough to discriminate between our proposed \texttt{DeepLens} model and a classical CNN model such as the one proposed in \citet{Petrillo2017}, despite the greater complexity of our model (46 vs.\ 7 layers). As an illustration, we show on \autoref{fig:resnet_vs_cnn} the ROC curve obtained by training and then evaluating a CNN following closely the description provided in \citet{Petrillo2017} on our set of simulations, along with the ROC curve obtained with our \texttt{DeepLens} model on the same training and testing set (single best epoch images). As can be seen, we find the two models to exhibit exactly the same performance when evaluated on our set of simulations, despite our model having outperformed most CNN based methods in the Euclid challenge (Metcalf et al. 2017, in prep.) and residual networks being known to significantly outperform CNNs in more complex image classification tasks \citep{He2015}. This result shows that a simpler CNN is already complex enough to capture all the variability present in our non-trivial yet simple simulations, and no significant gains are made by increasing the complexity of the model. More complex simulations including more variability and non-trivial contaminants would reveal the limits of simpler models, whereas the additional complexity of a deeper model would facilitate a better interpretation of complex galaxy images to find robust features for lens detection. This result also illustrates the important point that when using supervised machine learning approaches, most of the burden is shifted from the development of the method itself to the production of realistic simulations for training purposes.

It should also be noted that we limited the complexity of our fiducial model to 46 layers but it can easily be made much deeper, and similar architectures have been successfully trained up to a thousand layers \citep{He2016}. There is still room for substantially more model complexity to handle more realistic simulations and data.

In the simulations presented in this work, we have only used single band images, which demonstrates our model's ability to identify lenses from their morphologies alone. This is a valuable aspect of the method, especially for the Euclid survey, where lens searches will be conducted on the VIS single band images. However, colour information is also very helpful to identify strongly lensed systems and is already at the heart of some methods \citep[e.g.][]{Gavazzi2014}. As mentioned in the description of our architecture, our model can seamlessly handle multi-band images and we expect its performance to be significantly improved by the addition of colour information. 

Finally, we have  only considered in this work generic galaxy-galaxy lenses. However a number of particularly interesting science cases require exotic lens systems such as double source plane lenses \citep[e.g][]{Collett2014}, lensing catastrophes \citep{OrbanDeXivry2009} or lensed SuperNovae \cite[e.g.][]{Rodney2016}. However, these systems are much rarer and their specific detection is all the more challenging. In order to make our lens finder specifically sensitive to these rare cases, instead of a binary lens/non-lens classification, our method can be extended to a multi-class classification problem. A weighted cost function can also be used to further promote the purity of exotic lens candidates samples. 

\section{Conclusions}
\label{sec:conclusion}

In this work, we presented \texttt{CMU DeepLens}, a new strong gravitational lens finder based on the most recent advances in Deep Learning. Our fully automated method does not require any manual tuning or human intervention beyond the training step. This new class of machine learning techniques represents an exciting prospect for conducting large-scale lens searches, as they have been shown to surpass human abilities in similar image classification tasks. They have the potential to significantly cut down on the need for human visual inspection and make the search for strong lenses tractable in the era of deep yet wide-field surveys such as LSST, Euclid and WFIRST.  Being entirely automated, the selection function of such methods is also easier to quantify (given realistic simulations), which is a crucial point for precision cosmology \cite[e.g.][]{Collett2016a}.

We demonstrated on simple yet non-trivial strong lensing simulations of LSST $g$-band observations that our algorithm is extremely efficient, rejecting a vast majority of non-lenses while preserving a high level of completeness. As a point  of reference, we find that on single LSST exposures,  for a fiducial rejection rate of 99\%,  we still reach a completeness of 90\% for lenses  with Einstein radii larger  than $1.43 \arcsec$ and S/N of the lensed image larger than 20.  We also investigated the trade-off between better S/N at the cost of lower resolution by applying our lens finder to co-added image stacks, degraded to the worst seeing in the stack. We find very similar performance in this case compared  to single best seeing exposures. As a result we expect an optimal co-adding strategy to further improve on our results.

However, we note that these results are optimistic as our admittedly simple simulations do not include likely contaminants such as merging systems, spiral or ring galaxies. Our quoted completeness levels are therefore an optimistic estimate  of the performance of our model on real data. We further demonstrate that our simulations are too simple to discriminate between deep learning models of vastly different complexities, meaning that a conventional CNN model exhibits the same performance as our more advanced and far deeper residual network model. On more realistic simulations however, such as the ones used for the strong lens finding challenge (Metcalf et al.\ 2017, in prep.), our residual network architecture was found to benefit from this added complexity and to outperform more conventional CNNs. This result illustrates the need for realistic simulations when developing and applying supervised machine learning methods, thus shifting part of the effort from the model development to the production of realistic simulations. 

Finally, in the spirit of reproducible research, we make our code publicly available at \url{https://github.com/McWilliamsCenter/CMUDeepLens} .

The simulations used in this work will also be made available with an upcoming paper (Avestruz et al 2017, in prep).

\section*{Acknowledgements}
The authors would like to acknowledge Michelle Ntampaka, Hy Trac, Phil Marshall, Frederic Courbin, Remy Joseph, Sukhdeep Singh, Noble Kennamer, and Siyu He for useful remarks and discussions. We also would like to acknowledge the Euclid Strong Gravitational Lensing Challenge and the Bologna Lens Factory project for fostering the development of automated strong lens finders. This research is supported in part by DOE grant DESC0011114 and NSF grant IIS1563887.

%%%%%%%%%%%%%%%%%%%%%%%%%%%%%%%%%%%%%%%%%%%%%%%%%%



\bibliographystyle{mnras}
\bibliography{biblio} % if your bibtex file is called example.bib

\begin{thebibliography}{}
\makeatletter
\relax
\def\mn@urlcharsother{\let\do\@makeother \do\$\do\&\do\#\do\^\do\_\do\%\do\~}
\def\mn@doi{\begingroup\mn@urlcharsother \@ifnextchar [ {\mn@doi@}
  {\mn@doi@[]}}
\def\mn@doi@[#1]#2{\def\@tempa{#1}\ifx\@tempa\@empty \href
  {http://dx.doi.org/#2} {doi:#2}\else \href {http://dx.doi.org/#2} {#1}\fi
  \endgroup}
\def\mn@eprint#1#2{\mn@eprint@#1:#2::\@nil}
\def\mn@eprint@arXiv#1{\href {http://arxiv.org/abs/#1} {{\tt arXiv:#1}}}
\def\mn@eprint@dblp#1{\href {http://dblp.uni-trier.de/rec/bibtex/#1.xml}
  {dblp:#1}}
\def\mn@eprint@#1:#2:#3:#4\@nil{\def\@tempa {#1}\def\@tempb {#2}\def\@tempc
  {#3}\ifx \@tempc \@empty \let \@tempc \@tempb \let \@tempb \@tempa \fi \ifx
  \@tempb \@empty \def\@tempb {arXiv}\fi \@ifundefined
  {mn@eprint@\@tempb}{\@tempb:\@tempc}{\expandafter \expandafter \csname
  mn@eprint@\@tempb\endcsname \expandafter{\@tempc}}}

\bibitem[\protect\citeauthoryear{Alard}{Alard}{2006}]{Alard2006}
Alard C.,  2006, preprint (\mn@eprint {arXiv} {0606757})

\bibitem[\protect\citeauthoryear{Auger, Treu, Bolton, Gavazzi, Koopmans,
  Marshall, Moustakas  \& Burles}{Auger et~al.}{2010}]{Auger2010a}
Auger M.~W.,  Treu T.,  Bolton A.~S.,  Gavazzi R.,  Koopmans L. V.~E.,
  Marshall P.~J.,  Moustakas L.~A.,   Burles S.,  2010, \mn@doi [The
  Astrophysical Journal] {10.1088/0004-637X/724/1/511}, 724, 511

\bibitem[\protect\citeauthoryear{Barnab{\`{e}}, Czoske, Koopmans, Treu  \&
  Bolton}{Barnab{\`{e}} et~al.}{2011}]{Barnabe2011}
Barnab{\`{e}} M.,  Czoske O.,  Koopmans L. V.~E.,  Treu T.,   Bolton A.~S.,
  2011, \mn@doi [Monthly Notices of the Royal Astronomical Society]
  {10.1111/j.1365-2966.2011.18842.x}, 415, 2215

\bibitem[\protect\citeauthoryear{Bolton, Burles, Koopmans, Treu  \&
  Moustakas}{Bolton et~al.}{2006}]{Bolton2006}
Bolton A.~S.,  Burles S.,  Koopmans L. V.~E.,  Treu T.,   Moustakas L.~A.,
  2006, \mn@doi [The Astrophysical Journal] {10.1086/498884}, 638, 703

\bibitem[\protect\citeauthoryear{Bom, Makler, Albuquerque  \& Brandt}{Bom
  et~al.}{2017}]{Bom2016}
Bom C.~R.,  Makler M.,  Albuquerque M.~P.,   Brandt C.~H.,  2017, \mn@doi
  [Astronomy {\&} Astrophysics] {10.1051/0004-6361/201629159}, 597, A135

\bibitem[\protect\citeauthoryear{Bonvin et~al.,}{Bonvin
  et~al.}{2017}]{Bonvin2017}
Bonvin V.,  et~al., 2017, \mn@doi [Monthly Notices of the Royal Astronomical
  Society] {10.1093/mnras/stw3006}, 465, 4914

\bibitem[\protect\citeauthoryear{Brault \& Gavazzi}{Brault \&
  Gavazzi}{2015}]{Brault2015}
Brault F.,  Gavazzi R.,  2015, \mn@doi [Astronomy {\&} Astrophysics]
  {10.1051/0004-6361/201425275}, 577, A85

\bibitem[\protect\citeauthoryear{Cabanac et~al.,}{Cabanac
  et~al.}{2007}]{Cabanac2007}
Cabanac R.~A.,  et~al., 2007, \mn@doi [Astronomy and Astrophysics]
  {10.1051/0004-6361:20065810}, 461, 813

\bibitem[\protect\citeauthoryear{Cao, Biesiada, Gavazzi, Pi{\'{o}}rkowska  \&
  Zhu}{Cao et~al.}{2015}]{Cao2015}
Cao S.,  Biesiada M.,  Gavazzi R.,  Pi{\'{o}}rkowska A.,   Zhu Z.-H.,  2015,
  \mn@doi [The Astrophysical Journal] {10.1088/0004-637X/806/2/185}, 806, 185

\bibitem[\protect\citeauthoryear{Clevert, Unterthiner  \& Hochreiter}{Clevert
  et~al.}{2015}]{Clevert2015}
Clevert D.-A.,  Unterthiner T.,   Hochreiter S.,  2015, Under review of
  ICLR2016， 提出了ELU, pp 1--13

\bibitem[\protect\citeauthoryear{Collett}{Collett}{2015}]{Collett2015}
Collett T.~E.,  2015, \mn@doi [The Astrophysical Journal]
  {10.1088/0004-637X/811/1/20}, 811, 20

\bibitem[\protect\citeauthoryear{Collett \& Auger}{Collett \&
  Auger}{2014}]{Collett2014}
Collett T.~E.,  Auger M.~W.,  2014, \mn@doi [Monthly Notices of the Royal
  Astronomical Society] {10.1093/mnras/stu1190}, 443, 969

\bibitem[\protect\citeauthoryear{Collett \& Bacon}{Collett \&
  Bacon}{2016}]{Collett2016}
Collett T.~E.,  Bacon D.~J.,  2016, \mn@doi [Monthly Notices of the Royal
  Astronomical Society] {10.1093/mnras/stv2791}, 456, 2210

\bibitem[\protect\citeauthoryear{Collett \& Cunnington}{Collett \&
  Cunnington}{2016}]{Collett2016a}
Collett T.~E.,  Cunnington S.~D.,  2016, \mn@doi [Monthly Notices of the Royal
  Astronomical Society] {10.1093/mnras/stw1856}, 462, 3255

\bibitem[\protect\citeauthoryear{Connolly et~al.,}{Connolly
  et~al.}{2010}]{Connolly2010}
Connolly A.~J.,  et~al., 2010, \mn@doi [SPIE Astronomical Telescopes +
  Instrumentation] {10.1117/12.857819}, pp 77381O--77381O--10

\bibitem[\protect\citeauthoryear{Dieleman, Willett  \& Dambre}{Dieleman
  et~al.}{2015}]{Dieleman2015}
Dieleman S.,  Willett K.~W.,   Dambre J.,  2015, \mn@doi [Monthly Notices of
  the Royal Astronomical Society] {10.1093/mnras/stv632}, 450, 1441

\bibitem[\protect\citeauthoryear{Dye, Evans, Belokurov, Warren  \& Hewett}{Dye
  et~al.}{2008}]{Dye2008}
Dye S.,  Evans N.~W.,  Belokurov V.,  Warren S.~J.,   Hewett P.,  2008, \mn@doi
  [Monthly Notices of the Royal Astronomical Society]
  {10.1111/j.1365-2966.2008.13401.x}, 388, 384

\bibitem[\protect\citeauthoryear{Galametz et~al.,}{Galametz
  et~al.}{2013}]{Galametz2013}
Galametz A.,  et~al., 2013, \mn@doi [The Astrophysical Journal Supplement
  Series] {10.1088/0067-0049/206/2/10}, 206, 10

\bibitem[\protect\citeauthoryear{Gavazzi, Treu, Rhodes, Koopmans, Bolton,
  Burles, Massey  \& Moustakas}{Gavazzi et~al.}{2007}]{Gavazzi2007}
Gavazzi R.,  Treu T.,  Rhodes J.~D.,  Koopmans L. V.~E.,  Bolton A.~S.,  Burles
  S.,  Massey R.~J.,   Moustakas L.~A.,  2007, \mn@doi [The Astrophysical
  Journal] {10.1086/519237}, 667, 176

\bibitem[\protect\citeauthoryear{Gavazzi, Marshall, Treu  \&
  Sonnenfeld}{Gavazzi et~al.}{2014}]{Gavazzi2014}
Gavazzi R.,  Marshall P.~J.,  Treu T.,   Sonnenfeld A.,  2014, \mn@doi [The
  Astrophysical Journal] {10.1088/0004-637X/785/2/144}, 785, 144

\bibitem[\protect\citeauthoryear{Goodfellow, Bengio  \& Courville}{Goodfellow
  et~al.}{2016}]{Goodfellow2016}
Goodfellow I.,  Bengio Y.,   Courville A.,  2016, {Deep learning}.
MIT Press (\mn@eprint {arXiv} {arXiv:1312.6184v5})

\bibitem[\protect\citeauthoryear{Grogin et~al.,}{Grogin
  et~al.}{2011}]{Grogin2011}
Grogin N.~A.,  et~al., 2011, \mn@doi [The Astrophysical Journal Supplement
  Series December The Astrophysical Journal Supplement Series]
  {10.1088/0067-0049/197/2/35}, 197, 3535

\bibitem[\protect\citeauthoryear{He, Zhang, Ren  \& Sun}{He
  et~al.}{2015a}]{He2015}
He K.,  Zhang X.,  Ren S.,   Sun J.,  2015a, preprint, pp~-- (\mn@eprint
  {arXiv} {1512.03385})

\bibitem[\protect\citeauthoryear{He, Zhang, Ren  \& Sun}{He
  et~al.}{2015b}]{He2015a}
He K.,  Zhang X.,  Ren S.,   Sun J.,  2015b, in 2015 IEEE International
  Conference on Computer Vision (ICCV). IEEE, pp 1026--1034 (\mn@eprint {arXiv}
  {1502.01852}), \mn@doi{10.1109/ICCV.2015.123}, \url
  {http://arxiv.org/abs/1502.01852
  http://ieeexplore.ieee.org/document/7410480/}

\bibitem[\protect\citeauthoryear{He, Zhang, Ren  \& Sun}{He
  et~al.}{2016}]{He2016}
He K.,  Zhang X.,  Ren S.,   Sun J.,  2016, preprint, pp 1--15 (\mn@eprint
  {arXiv} {1603.05027})

\bibitem[\protect\citeauthoryear{Hinton, Srivastava, Krizhevsky, Sutskever  \&
  Salakhutdinov}{Hinton et~al.}{2012}]{Hinton2012}
Hinton G.~E.,  Srivastava N.,  Krizhevsky A.,  Sutskever I.,   Salakhutdinov
  R.~R.,  2012, preprint, pp 1--18 (\mn@eprint {arXiv} {1207.0580})

\bibitem[\protect\citeauthoryear{Hornik, Stinchcombe  \& White}{Hornik
  et~al.}{1989}]{Hornik1989}
Hornik K.,  Stinchcombe M.,   White H.,  1989, \mn@doi [Neural Networks]
  {10.1016/0893-6080(89)90020-8}, 2, 359

\bibitem[\protect\citeauthoryear{Hoyle}{Hoyle}{2016}]{Hoyle2016}
Hoyle B.,  2016, \mn@doi [Astronomy and Computing]
  {10.1016/j.ascom.2016.03.006}, 16, 34

\bibitem[\protect\citeauthoryear{Joseph et~al.,}{Joseph
  et~al.}{2014}]{Joseph2014}
Joseph R.,  et~al., 2014, \mn@doi [Astronomy {\&} Astrophysics]
  {10.1051/0004-6361/201423365}, 566, A63

\bibitem[\protect\citeauthoryear{Kim \& Brunner}{Kim \&
  Brunner}{2017}]{Kim2017}
Kim E.~J.,  Brunner R.~J.,  2017, \mn@doi [Monthly Notices of the Royal
  Astronomical Society] {10.1093/mnras/stw2672}, 464, 4463

\bibitem[\protect\citeauthoryear{Kingma \& Ba}{Kingma \& Ba}{2015}]{Kingma2015}
Kingma D.~P.,  Ba J.~L.,  2015, International Conference on Learning
  Representations 2015, pp 1--15

\bibitem[\protect\citeauthoryear{Koekemoer et~al.,}{Koekemoer
  et~al.}{2011}]{Koekemoer2011}
Koekemoer A.~M.,  et~al., 2011, \mn@doi [The Astrophysical Journal Supplement
  Series December The Astrophysical Journal Supplement Series]
  {10.1088/0067-0049/197/2/36}, 197

\bibitem[\protect\citeauthoryear{Koopmans, Treu, Bolton, Burles  \&
  Moustakas}{Koopmans et~al.}{2006}]{Koopmans2006}
Koopmans L. V.~E.,  Treu T.,  Bolton A.~S.,  Burles S.,   Moustakas L.~A.,
  2006, \mn@doi [The Astrophysical Journal] {10.1086/505696}, 649, 599

\bibitem[\protect\citeauthoryear{Krizhevsky, Sutskever  \& Hinton}{Krizhevsky
  et~al.}{2012}]{Krizhevsky2012}
Krizhevsky A.,  Sutskever I.,   Hinton G.~E.,  2012, \mn@doi [Advances In
  Neural Information Processing Systems]
  {http://dx.doi.org/10.1016/j.protcy.2014.09.007}, pp~1--9

\bibitem[\protect\citeauthoryear{Kubo \& Dell'Antonio}{Kubo \&
  Dell'Antonio}{2008}]{Kubo2008}
Kubo J.~M.,  Dell'Antonio I.~P.,  2008, \mn@doi [Monthly Notices of the Royal
  Astronomical Society] {10.1111/j.1365-2966.2008.12880.x}, 385, 918

\bibitem[\protect\citeauthoryear{{LSST Science Collaboration} et~al.,}{{LSST
  Science Collaboration} et~al.}{2009}]{LSST2009}
{LSST Science Collaboration} et~al., 2009, preprint, p. arXiv:0912.0201
  (\mn@eprint {arXiv} {0912.0201})

\bibitem[\protect\citeauthoryear{Laureijs et~al.,}{Laureijs
  et~al.}{2011}]{Laureijs2011}
Laureijs R.,  et~al., 2011, preprint, p. arXiv:1110.3193 (\mn@eprint {arXiv}
  {1110.3193})

\bibitem[\protect\citeauthoryear{LeCun, Bengio  \& Hinton}{LeCun
  et~al.}{2015}]{LeCun2015}
LeCun Y.,  Bengio Y.,   Hinton G.,  2015, \mn@doi [Nature]
  {10.1038/nature14539}, 521, 436

\bibitem[\protect\citeauthoryear{Lecun, Bottou, Bengio  \& Haffner}{Lecun
  et~al.}{1998}]{Lecun1998}
Lecun Y.,  Bottou L.,  Bengio Y.,   Haffner P.,  1998, \mn@doi [Proceedings of
  the IEEE] {10.1109/5.726791}, 86, 2278

\bibitem[\protect\citeauthoryear{Li, Gladders, Rangel, Florian, Bleem,
  Heitmann, Habib  \& Fasel}{Li et~al.}{2016}]{Li2016}
Li N.,  Gladders M.~D.,  Rangel E.~M.,  Florian M.~K.,  Bleem L.~E.,  Heitmann
  K.,  Habib S.,   Fasel P.,  2016, \mn@doi [The Astrophysical Journal]
  {10.3847/0004-637X/828/1/54}, 828, 54

\bibitem[\protect\citeauthoryear{Marshall, Hogg, Moustakas, Fassnacht,
  Brada{\v{c}}, Schrabback  \& Blandford}{Marshall et~al.}{2009}]{Marshall2009}
Marshall P.~J.,  Hogg D.~W.,  Moustakas L.~A.,  Fassnacht C.~D.,  Brada{\v{c}}
  M.,  Schrabback T.,   Blandford R.~D.,  2009, \mn@doi [The Astrophysical
  Journal] {10.1088/0004-637X/694/2/924}, 694, 924

\bibitem[\protect\citeauthoryear{Marshall et~al.,}{Marshall
  et~al.}{2015}]{Marshall2015}
Marshall P.~J.,  et~al., 2015, \mn@doi [Monthly Notices of the Royal
  Astronomical Society] {10.1093/mnras/stv2009}, 455, 1171

\bibitem[\protect\citeauthoryear{More, Cabanac, More, Alard, Limousin, Kneib,
  Gavazzi  \& Motta}{More et~al.}{2012}]{More2012}
More A.,  Cabanac R.,  More S.,  Alard C.,  Limousin M.,  Kneib J.-P.,  Gavazzi
  R.,   Motta V.,  2012, \mn@doi [The Astrophysical Journal]
  {10.1088/0004-637X/749/1/38}, 749, 38

\bibitem[\protect\citeauthoryear{More et~al.,}{More et~al.}{2015}]{More2015}
More A.,  et~al., 2015, \mn@doi [Monthly Notices of the Royal Astronomical
  Society] {10.1093/mnras/stv1965}, 455, 1191

\bibitem[\protect\citeauthoryear{Nair \& Hinton}{Nair \&
  Hinton}{2010}]{Nair2010}
Nair V.,  Hinton G.~E.,  2010, \mn@doi [Proceedings of the 27th International
  Conference on Machine Learning] {10.1.1.165.6419}, pp 807--814

\bibitem[\protect\citeauthoryear{Oguri \& Marshall}{Oguri \&
  Marshall}{2010}]{Oguri2010}
Oguri M.,  Marshall P.~J.,  2010, \mn@doi [Monthly Notices of the Royal
  Astronomical Society] {10.1111/j.1365-2966.2010.16639.x}, 405, 2579

\bibitem[\protect\citeauthoryear{{Orban De Xivry} \& Marshall}{{Orban De Xivry}
  \& Marshall}{2009}]{OrbanDeXivry2009}
{Orban De Xivry} G.,  Marshall P.,  2009, \mn@doi [Monthly Notices of the Royal
  Astronomical Society] {10.1111/j.1365-2966.2009.14925.x}, 399, 2

\bibitem[\protect\citeauthoryear{Petrillo et~al.,}{Petrillo
  et~al.}{2017}]{Petrillo2017}
Petrillo C.~E.,  et~al., 2017, preprint, 19, 1 (\mn@eprint {arXiv}
  {1702.07675})

\bibitem[\protect\citeauthoryear{Ravanbakhsh, Lanusse, Mandelbaum, Schneider
  \& Poczos}{Ravanbakhsh et~al.}{2016}]{Ravanbakhsh2016}
Ravanbakhsh S.,  Lanusse F.,  Mandelbaum R.,  Schneider J.,   Poczos B.,  2016,
  preprint, pp~1--7 (\mn@eprint {arXiv} {1609.05796})

\bibitem[\protect\citeauthoryear{Refsdal}{Refsdal}{1964}]{Refsdal1964}
Refsdal S.,  1964, \mn@doi [Monthly Notices of the Royal Astronomical Society]
  {10.1093/mnras/128.4.307}, 128, 307

\bibitem[\protect\citeauthoryear{Rodney et~al.,}{Rodney
  et~al.}{2016}]{Rodney2016}
Rodney S.~A.,  et~al., 2016, \mn@doi [The Astrophysical Journal]
  {10.3847/0004-637X/820/1/50}, 820, 50

\bibitem[\protect\citeauthoryear{Rumelhart, Hinton  \& Williams}{Rumelhart
  et~al.}{1986}]{Rumelhart1986}
Rumelhart D.~E.,  Hinton G.~E.,   Williams R.~J.,  1986, \mn@doi [Nature]
  {10.1038/323533a0}, 323, 533

\bibitem[\protect\citeauthoryear{Seidel \& Bartelmann}{Seidel \&
  Bartelmann}{2007}]{Seidel2007}
Seidel G.,  Bartelmann M.,  2007, \mn@doi [Astronomy and Astrophysics]
  {10.1051/0004-6361:20066097}, 472, 341

\bibitem[\protect\citeauthoryear{Spergel et~al.,}{Spergel
  et~al.}{2015}]{Spergel2015}
Spergel D.,  et~al., 2015, preprint (\mn@eprint {arXiv} {1503.03757})

\bibitem[\protect\citeauthoryear{Srivastava, Hinton, Krizhevsky, Sutskever  \&
  Salakhutdinov}{Srivastava et~al.}{2014}]{Srivastava2014}
Srivastava N.,  Hinton G.,  Krizhevsky A.,  Sutskever I.,   Salakhutdinov R.,
  2014, \mn@doi [Journal of Machine Learning Research] {10.1214/12-AOS1000},
  15, 1929

\bibitem[\protect\citeauthoryear{Suyu, Marshall, Auger, Hilbert, Blandford,
  Koopmans, Fassnacht  \& Treu}{Suyu et~al.}{2010}]{Suyu2010}
Suyu S.~H.,  Marshall P.~J.,  Auger M.~W.,  Hilbert S.,  Blandford R.~D.,
  Koopmans L. V.~E.,  Fassnacht C.~D.,   Treu T.,  2010, \mn@doi [The
  Astrophysical Journal] {10.1088/0004-637X/711/1/201}, 711, 201

\bibitem[\protect\citeauthoryear{Szegedy et~al.,}{Szegedy
  et~al.}{2015}]{Szegedy2015}
Szegedy C.,  et~al., 2015, \mn@doi [Proceedings of the IEEE Computer Society
  Conference on Computer Vision and Pattern Recognition]
  {10.1109/CVPR.2015.7298594}, 07-12-June, 1

\bibitem[\protect\citeauthoryear{Treu}{Treu}{2010}]{Treu2010}
Treu T.,  2010, \mn@doi [Annual Review of Astronomy and Astrophysics]
  {10.1146/annurev-astro-081309-130924}, 48, 87

\bibitem[\protect\citeauthoryear{Zahid, Damjanov, Geller  \&
  Chilingarian}{Zahid et~al.}{2015}]{Zahid2015}
Zahid H.~J.,  Damjanov I.,  Geller M.~J.,   Chilingarian I.,  2015, \mn@doi
  [The Astrophysical Journal] {10.1088/0004-637X/806/1/122}, 806, 122

\bibitem[\protect\citeauthoryear{de Jong et~al.,}{de~Jong
  et~al.}{2015}]{DeJong2015}
de Jong J. T.~A.,  et~al., 2015, \mn@doi [Astronomy {\&} Astrophysics]
  {10.1051/0004-6361/201526601}, 582, A62

\makeatother
\end{thebibliography}

%%%%%%%%%%%%%%%%%%%%%%%%%%%%%%%%%%%%%%%%%%%%%%%%%%

% Don't change these lines
\bsp	% typesetting comment
\label{lastpage}
\end{document}